\documentclass[onecolumn,authoryear]{els-mrw}

\usepackage{amsmath,amssymb,amsfonts,amsthm,makeidx,graphicx}
\usepackage{txfonts}
\usepackage{helvet}
\usepackage[colorlinks=true,citecolor=blue, urlcolor=blue, ]{hyperref}
\usepackage[normalem]{ulem}
\usepackage{xcolor}
\usepackage{wrapfig}

\newcommand{\presecContentsEntry}[1]{\noindent\nameref{#1} \hfill \pageref{#1} \newline}
\newcommand{\secContentsEntry}[1]{\noindent\ref{#1}. \nameref{#1} \hfill \pageref{#1} \newline}
\newcommand{\subsecContentsEntry}[1]{\forceindent\ref{#1}. \nameref{#1} \hfill \pageref{#1}\newline}
\newcommand{\forceindent}{\leavevmode{\parindent=2em\indent}}

\newcommand{\horizontalLine}[1]{\noindent\textcolor{darkgray}{\rule{\textwidth}{1pt} }}

\newcommand{\ha}{H$\alpha$}

\usepackage{soul}

\def\pasa{PASA}
\def\physrep{PhR}
\def\ssr{Space Sci.~Rev.}
\def\baas{Bull.~Am.~Astron.~Soc.}
\def\aapr{A\&ARv}
\def\aaps{A\&AS}
\def\aj{AJ}
\def\apj{ApJ}

\def\apjl{ApJL}
\def\apjs{ApJS}

\def\aap{A\&A}
\def\araa{ARAA}
\def\mnras{MNRAS}

\def\prd{PRD}

\def\pasp{PASP}

\def\nat{Nature}

\DeclareRobustCommand{\ion}[2]{%
\relax\ifmmode
\ifx\testbx\f@series
{\mathbf{#1\,\mathsc{#2}}}\else
{\mathrm{#1\,\mathsc{#2}}}\fi
\else\textup{#1\,{\mdseries\textsc{#2}}}%
\fi}
\newcommand{\msun}{\mbox{${\rm M}_\odot$}}
\newcommand{\mstar}{\mbox{${M}_{\rm star}$}}
\newcommand{\cmjj}{\mbox{${\rm cm^{-2}}$}}
\newcommand{\cmjjj}{\mbox{${\rm cm^{-3}}$}}
\newcommand{\lya}{\mbox{${\rm Ly}\alpha$}}

\newcommand{\kms}{\mbox{${\rm km\,s}^{-1}$}}

\begin{document}

\graphicspath{{./figures/}}

\chapter{The Circumgalactic Medium}\label{ch:circumgalactic-meddium}

\author[1]{Hsiao-Wen Chen}
\author[2]{Fakhri S.~Zahedy}

\address[1]{\orgname{The University of Chicago}, \orgdiv{Department of
    Astronomy \& Astrophysics}, \orgaddress{5640 S.\ Ellis Ave., Chicago, IL 60637, USA}}

\address[2]{\orgname{University of North Texas}, \orgdiv{Department of Physics}, \orgaddress{210 Avenue A, Denton, TX 76201, USA}}


\maketitle

\begin{abstract}[Abstract]
Galaxies are part of a vast cosmic ecosystem, embedded in an extensive gaseous reservoir that regulates their growth by providing the necessary fuel for star formation while preserving a fossil record of past interactions, outflows, and feedback processes. The circumgalactic medium (CGM) contains multiphase gas spanning a broad dynamic range in spatial scale, density, and temperature, with its thermodynamic and chemical properties deeply linked to the star formation histories of galaxies. As a rich laboratory for studying gas physics, the CGM offers unique insights into the processes governing gas cooling, heating, and material transfer between galaxies and their surroundings. Chemical tagging, based on the relative abundances of multiple elements, serves as a powerful timing tool to trace the origin of the gas and connect the stars in the interstellar medium (ISM) to the diffuse CGM. Developing a complete understanding of the CGM and its cosmic evolution requires multi-wavelength observational tools, ranging from X-ray, UV/optical, and sub-mm to radio, to probe this diffuse gas in both absorption and emission.
\end{abstract}


\vspace{5mm}

\begin{keywords}
   galaxy halos, cosmic ecosystems, quasar absorption lines, the baryon cycle
\end{keywords}

\vspace{5mm}

\horizontalLine

\noindent \textbf{Contents}

\presecContentsEntry{presec:glossary}
\presecContentsEntry{presec:nomenclature}
\secContentsEntry{sec:intro}
\subsecContentsEntry{sec:eco}
\subsecContentsEntry{sec:history}
\subsecContentsEntry{sec:questions}
\secContentsEntry{sec:phases}
\secContentsEntry{sec:methods}
\subsecContentsEntry{sec:emission}
\subsecContentsEntry{sec:abs}
\secContentsEntry{sec:empirical}
\subsecContentsEntry{sec:kappa}
\subsecContentsEntry{sec:resolve}
\subsecContentsEntry{sec:thermo}
\subsecContentsEntry{sec:chemical}
\secContentsEntry{sec:future}
\subsecContentsEntry{sec:multisightlines}
\subsecContentsEntry{sec:others}

\horizontalLine


\begin{BoxTypeA}[]{Key points and learning objectives}
\begin{itemize}
\item Learn about the intricate connection between the circumgalactic medium (CGM) and galaxy formation and evolution. 
\item Develop an understanding of the complex multiphase and multiscale physics that shapes the diffuse cosmic gas. 
\item Apply chemical tagging as a timing clock for establishing chemical enrichment history 
\item Identify the tools needed to establish a holistic understanding of the diffuse cosmic gas.  
\end{itemize}
\end{BoxTypeA}

\clearpage

\begin{glossary}[Glossary] 
\label{presec:glossary}
  \term{pc.} One parsec equals to $3.086\times 10^{18}$ cm, which is roughly 3.26 light years; $1\,{\rm kpc}=1000\,{\rm pc}$; $1\,{\rm Mpc}=1000,000\,{\rm pc}$. \\
  \term{Doppler parameter.} The spectral line width of an astronomical object, caused by the internal velocity dispersion along the line of sight. \\
  \term{metal.} elements heavier than hydrogen and helium.\\
  \term{Voigt profile.} A mathematical representation of a spectral line, involving a convolution of a Lorentzian distribution and a Gaussian function to effectively capture the intrinsic and thermal broadening that shape spectral lines from astronomical objects. \\
  \term{resonant line.} A spectral feature produced by an electron moving between the ground state and the first excited state of an atom. \\
  \term{forbidden line.} A spectral feature with a very low transition probability due to the selection rules governed by the principles of quantum mechanics. These transitions are commonly seen in low-density astrophysical environments. \\
  \term{nebular line.} An emission line commonly seen in ionized gas in \ion{H}{II} regions or planetary nebulae. It is produced by electrons moving from excited states to lower orbits. \\
  \term{$\alpha$-element.} Elements formed by $\alpha$-capture process.  \\
\end{glossary}

\begin{glossary}[Nomenclature] 
\label{presec:nomenclature}

  \begin{tabular}{@{}lp{34pc}@{}}
    AGN & active galactic nucleus powered by a supermassive black hole in the center of a galaxy\\
    BPT & Baldwin, Phillps, \& Terlevich diagram \\
    CGM & circumgalactic medium \\
    \ion{H}{I} & neutral hydrogen \\
    \ion{H}{II} & ionized hydrogen, one electron removed \\
    ICM & intracluster medium \\
    IGM & intergalactic medium \\
    IFS & integral field spectrograph \\
    ISM & interstellar medium \\
    QSO & quasi-stellar object \\
    quasar & the most luminous class of AGN \\
    SMBH & supermassive black hole \\
    VSF & velocity structure function \\
    WHIM & warm-hot ionized medium \\
    $z$ & redshift, a dimensionless measure of the recession velocity of an external object \\
    $Z$ & Metallicity, that is fraction of elements heavier than helium\\
\end{tabular}
\end{glossary}

\section{Introduction} \label{sec:intro}




Modern cosmology represents a major triumph of the scientific method. Today's standard model of cosmology, 
also known as the $\Lambda$CDM (Lambda Cold Dark Matter) model, can explain and connect the observed large-scale 
structures of the Universe today, based on large spectroscopic surveys of galaxies \citep[e.g,][]{York:2000}, to the primordial density and 
temperature fluctuations in the early Universe observed as anisotropies in the power spectrum of the Cosmic Microwave 
Background \citep[see a review by][]{Bullock:2017}. 
Despite $\Lambda$CDM cosmology's significant success in explaining the formation of large-scale structures ($\gtrsim10$ Mpc scales) and 
their evolution over almost 14 Gyr of cosmic time, major questions remain on physical processes acting on smaller 
scales. This smaller scale ($\lesssim1$ Mpc) is the realm of the astrophysics of galaxy formation and evolution.

\subsection{Cosmic ecosystems}
\label{sec:eco}

Galaxies are the fundamental building blocks of our Universe, hosting various processes, with star formation being one of their primary activities. 
However, a puzzle emerges: star-forming galaxies across all masses lack enough material in their interstellar medium (ISM) to sustain star formation for a period comparable to their current age. The typical gas depletion timescale for these galaxies, covering a wide range of masses, is less than a few Gyr.  
\citep[e.g.,][]{Bigiel:2011, Kennicutt:2012}. The solution lies in the understanding that galaxies are not closed systems---they rely on an external supply of gas to replenish their ISM and sustain star formation over the lifetime of the universe. 

\begin{figure}[tb]
  \centering
  \includegraphics[width=0.95\textwidth]{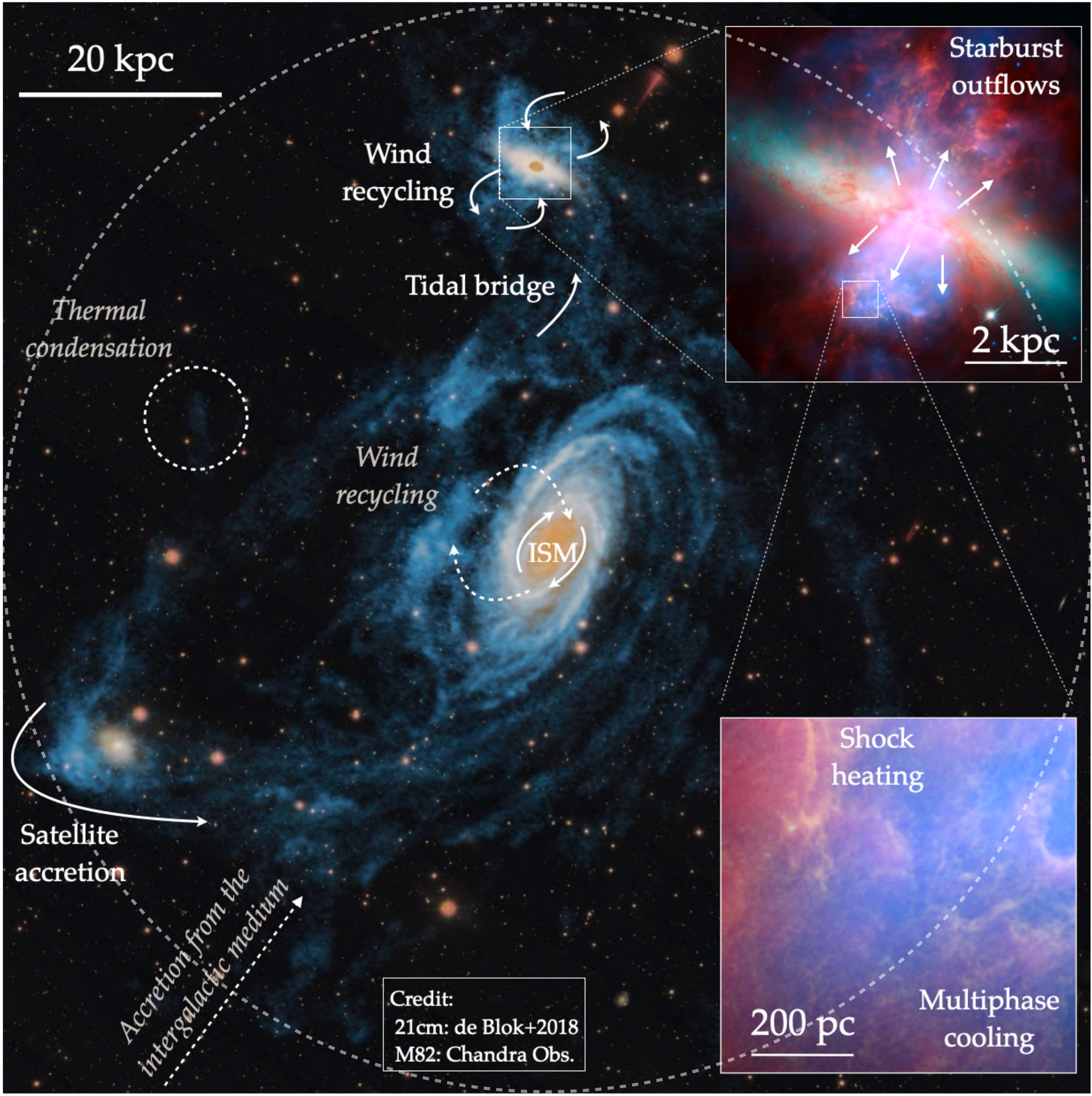} 
  \caption{Composite images of regions around the M81 group, revealing the intricate multi-scale physical processes that govern the cycling of gas between the multiphase circumgalactic medium (CGM), stars, and galaxies within the cosmic ecosystems.  \ion{H}{I} 21~cm images covering a region of $\approx\,120$ kpc on a side centered at M81 are overlaid on top of optical images obtained from the Sloan Digital Sky Survey \citep[SDSS;][]{York:2000}. This multi-wavelength composite image \citep[adapted from][]{deblok:2018} showcases widespread diffuse gaseous streams uncovered by \ion{H}{I} 21~cm observations connecting between what appear to be isolated galaxies in the optical images. The rich morphology in the diffuse \ion{H}{I} emission enables identifications of key processes such as wind recycling, satellite accretion, tidal stripping, and possibly thermal condensation and accretion from the IGM in the broader CGM around M81. In addition, the inset in the upper-right corner displays a zoomed-in composite image of the proto-typical starburst galaxy, M82, adapted from the Chandra Observatory imaging archive).  Combining X-ray and optical H$\alpha$ images reveals powerful, multiphase starburst-driven outflowing materials reaching beyond 6 kpc in the projected distance along the polar axis. At the same time, a second inset in the lower-right corner highlights exquisite details of shock heating and multiphase cooling regions on still smaller scales of $\sim 10$ pc. }
  \label{fig:cgmim}
\end{figure}

Over the past few decades, observations have established that most galaxies, from dwarf galaxies in our cosmic neighborhood to Milky-Way analogs and giant ellipticals in the distant Universe, are surrounded by spatially extended gaseous envelope known as the circumgalactic medium (CGM). Within this gaseous envelope is the stage on which the complex and multi-scale physical processes of the baryon cycle, the two-way exchange of baryonic matter between galaxies and their environments, play out. As illustrated in Fig.\ \ref{fig:cgmim}, a composite image from combining multi-wavelength observations, galaxies accrete gas from the surrounding intergalactic medium (IGM), which can then replenish the ISM and fuel star formation. In turn, vigorous star formation activity can trigger outflows, powered by supernova explosions, that expel gas and heavy elements out of galaxies. The fate of such expelled material varies: some will eventually fall back onto galaxies if their potential wells are deep enough. Others may escape the gravitational pull of the galaxies and enrich the IGM instead. In addition, gravitational interactions between galaxies and the hydrodynamic effects of satellite galaxies moving through a hot halo can lead to tidal stripping and ram-pressure stripping of gas from satellite members within galaxy groups \citep[e.g.,][]{Gunn:1972}. 

There is now growing recognition that the CGM plays a key role in galaxy evolution \citep[][]{Decadal2020}. With the understanding that galaxies are not isolated island universes, but instead integral parts of a cosmic ecosystem, then studying the CGM is therefore akin to developing the science of economics for galaxies. Like economics, it promotes a fuller understanding of galaxy evolution by investigating how galaxies obtain, recycle, and exchange resources (gas) and goods (heavy elements) with their environment, and how these processes impact the course of galaxy evolution. 

\subsection{A historical overview of circumgalactic research}
\label{sec:history}
In the mid-20th century, Guido M\"{u}nch identified gas clouds at significant heights ($\sim\!0.5$--1 kpc) above 
the Galactic disk. His observations were based on absorption lines from neutral sodium (\ion{Na}{I}) and singly ionized calcium (\ion{Ca}{II}) ions 
in the spectra of blue, hot stars at high Galactic latitudes.  These findings were first noted in \cite{Spitzer:1956} and later formally published by \cite{Munch:1961}. The presence of these neutral and low-ionization species implies that these clouds are cool ($T\lesssim 10^4$ K) and dense ($n_\mathrm{H}\sim 10$ hydrogen atoms \cmjjj). Furthermore, their observability requires that these clouds are pressure confined, which then led \citet[][]{Spitzer:1956} to theorize the existence of a hot ($T\!\sim\!10^6$ K), diffuse ($n_\mathrm{H}\!\sim\!10^{-4}$ \cmjjj), and volume-filling gaseous corona that surrounds the Milky Way.
 
M\"{u}nch's discovery illustrates the power of absorption-line spectroscopy for detecting and characterizing intervening gas seen along background lines of sight (see Section \ref{sec:methods} and Fig.\ \ref{fig:qal} below). This observational method would experience its first revolution sometime in the following decade, with the discovery of a class of blue point sources now known as quasars or quasi-stellar objects (QSOs). The extragalactic nature and vast, cosmological distances to these QSOs were almost immediately recognized \citep[][]{Schmidt:1963,Oke:1963}, and many investigators started exploring the possibility of using background QSOs to detect tenuous gas outside the Galaxy \citep[][]{Bahcall:1965, Bahcall:1966, Rees:1967}. 
Within only a few years, various authors had established that intervening absorption lines are a ubiquitous feature in the spectra of QSOs 
\citep[e.g.,][]{Burbidge:1966, Greenstein:1967, Bahcall:1967}. 

At the end of 1960s, \citet[][]{Bahcall:1969} postulated that a majority of these absorption lines are produced by diffuse gas in the extended gaseous halos surrounding galaxies, as predicted by the earlier \citet[][]{Spitzer:1956} study. However, it would be another decade before the first associations between galaxies and QSO absorption-line systems could be made, when several works identified
galaxies with redshifts coincident with metal absorption species, specifically the \ion{Ca}{II}\,$\lambda\lambda 3934, 3969$ and \ion{Mg}{II}\,$\lambda\lambda 2796, 2803$ doublets previously detected in QSO spectra \citep[e.g.,][]{Boksenberg:1978, Boksenberg:1980, Bergeron:1986}. A common feature of these findings were the large projected distances between the QSO lines of sight and the foreground galaxies associated with the absorption, far above the disk size and consistent with the extended halo physical picture of \citet[][]{Bahcall:1969}. 

In the few decades following these pioneering studies, observational studies of the CGM have continued to rely largely on the absorption lines imprinted by the intervening CGM gas in QSO spectra. In particular, the advent of the {\it Hubble Space Telescope} ({\it HST}) and large ground-based telescopes in the 1990s and early 2000s triggered a second renaissance of the field by allowing systematic studies to characterize the extended gaseous halos of galaxies using large samples ($\gtrsim$ dozens) 
of background QSO-foreground galaxy pairs  \citep[e.g.,][]{Morris:1993, Lanzetta:1995, Steidel:1997, Chen:1998, Chen:2001}, a trend that has continued to the present day. 

\subsection{The CGM and key questions in galaxy evolution}
\label{sec:questions}
In the $\Lambda$CDM paradigm, linear matter overdensities seeded during the early Universe grow over time until they reach a point where self-gravity begins to dominate, and they collapse to form dark-matter halos. Because baryonic matter is subdominant to dark matter in mass, the fate of these baryons is to be accreted into the gravitational potential wells of dark matter halos. The accreted baryons cool radiatively and condense over time as they fall deeper into the center of a dark-matter halo \citep[][]{Rees:1977, White:1978}, eventually forming dense, molecular gas complexes stars are born out of. The primary objective of galaxy evolution studies is to explain how from a seemingly simple beginning came the incredibly diverse population of galaxies seen across cosmic time. 
Here we outline several major problems in galaxy evolution studies and motivate how the CGM is linked with each one.


1. \textbf{Why have galaxies turned so little of their baryons into stars?} In the present day, galaxies across all masses harbor only a small fraction of their dark matter halo's cosmological budget of baryons in stars and ISM gas. While dark-matter halos hosting Milky-Way sized galaxies appear to the most efficient in turning their gas to stars, even then fewer than 20\% of their budgeted baryons are locked in stars. Lower mass halos hosting dwarf galaxies and massive halos hosting giant elliptical galaxies have done far worse, with fewer than $\lesssim5-10$ \% of their baryons converted into stars \citep[e.g.,][]{Kravtsov:2018}. Such mass dependence in a dark matter halo's global star formation efficiency suggests that multiple physical processes are at play across different mass regimes to regulate star-formation processes. The CGM is an excellent venue to search for evidence of the physical processes responsible for the regulation of star formation. 

2. \textbf{Where are the baryons and how did they get so far away from galaxies?} If a great majority of baryons are not locked in stars within galaxies, where are they? The first possible answer is that most baryons have never been incorporated into galaxy halos to begin with, and they remain in 
the IGM. While observations of the Lyman-alpha (Ly$\alpha$) forest indicate that this is the case at high redshifts \citep[$z>2$, e.g., ][]{Rauch:1998}, the cool IGM's share of the cosmic baryon budget appears to be a factor of several times lower in the present day. Another possibility is that a large fraction of baryons were once accreted into dark matter halos, but they had since been heated and possibly expelled out of halos due to energetic feedback processes from galaxies, and they now reside in a warm-hot ionized phase \citep[WHIM; e.g.,][]{Bregman:2007, McQuinn:2016}. However, detecting WHIM is extremely challenging due to the expected low densities and high temperatures ($T\sim10^{5-7}\,$K; $n_\mathrm{H}<10^{-4}\,$\cmjjj), and empirical constraints on their baryonic content remain highly uncertain to date. The final scenario is that baryons reside within dark matter halos, but unlike ISM gas they are too diffuse and therefore difficult to detect in emission. Characterizing the physical properties of the CGM and accounting for its various gas phases and how they relate to the baryon budget is therefore of great importance to solving this problem and distinguishing between the three scenarios above.  

3.  \textbf{Why do some galaxies stop forming stars altogether and stay that way?} Among different populations of galaxies, massive quiescent galaxies present
some of the most perplexing problems in galaxy evolution studies. These massive galaxies, with total stellar masses of $M_\mathrm{star}\gtrsim10^{11}M_\odot$ and 
residing in dark-matter halos with of $M_\mathrm{h}\gtrsim10^{13}M_\odot$, are dominated by old ($\gtrsim 5$ Gyr) stellar populations \citep[e.g.,][]{Tojeiro:2011}.
Unlike lower-mass galaxies like our Milky Way, these massive galaxies show little ongoing star formation and appear to have ceased forming stars for many Gyr \citep[e.g.,][]{Ferreras:2009}. A natural expectation is that these galaxies lack the cool gas needed to replenish their ISM and restart star formation. Searching and characterizing the cool CGM of these massive quiescent galaxies is, therefore, necessary to advance our understanding of massive galaxy evolution. 


\section{The different phases of the CGM} \label{sec:phases}

The multiphase nature of the CGM is reflected by the co-presence of gas at different temperatures, densities, and ionization states in both observations and simulation studies.  Observationally, multi-wavelength images of the Milky Way and nearby galaxies have revealed complex mixtures of different gas phases \citep[e.g.,][]{Putman:2012}. An example of the M81 group with a total mass of $M_h\!\approx\!10^{12}\,\msun$ \citep[e.g.,][]{Karachentsev:2006} is displayed in Fig.\ \ref{fig:cgmim}. Composites of radio 21~cm observations (targeting neutral hydrogen gas), optical narrow-band images (targeting nebular lines from ionized gas), and X-ray images (targeting hot plasma) unveil the intricate structures of cold neutral gaseous streams of $T\!\lesssim\!1000$ K connecting between member galaxies of the group across $\sim\!100$ kpc in the projected distance.  These are in addition to 
detailed multiphase gaseous structures on scales of $\sim\!{\rm pc}$ in the biconical outflows from M82. This multiphase gas fills up the immense space between luminous stellar bodies.  What is missing in this image is the ambient hot atmosphere too faint to be detected using existing X-ray telescopes, but routinely observed around more massive galaxies and galaxy clusters of mass exceeding $M_h\!\approx\!10^{14}\,\msun$ in the nearby universe (e.g., \citealt{Sun:2009}, \citealt{OSullivan:2017}; see also \citealt{Donahue2022} for a recent review).  In halos hosting galaxy groups of $M_h\!\approx\!10^{13}\,\msun$, ionized hot plasma has been detected through absorption features of high-ionization species, such as \ion{O}{VI} and \ion{O}{VII}, which correspond to oxygen atoms that have been ionized five and six times, respectively \citep[e.g.,][]{Mulchaey:1996, Fang:2002} and the Sunyaev-Zel'dovich signals against the cosmic microwave background \citep[e.g.,][]{Carlstrom:2002, Hand:2011, Pratt:2021, Hadzhiyska:2024}.

\begin{figure}[t]
  \centering
  \includegraphics[width=0.97\textwidth]{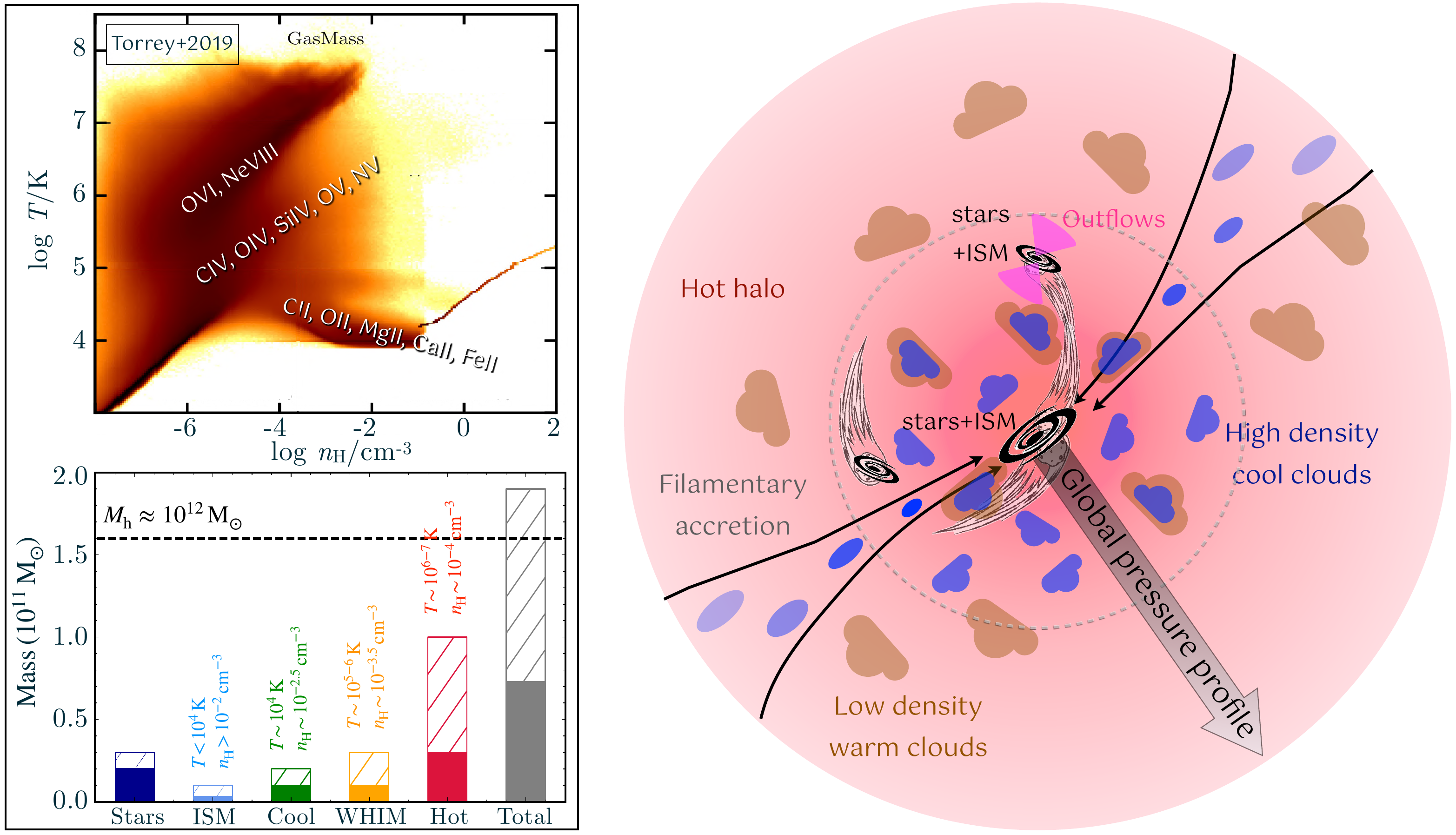}
  \caption{The galactic atmosphere, encompassing the multiphase CGM, spans a broad range in temperature and density.  Cosmological simulations have shown that the gas content of the Universe can be broadly classified into four different regimes, from cool, photoionized IGM of temperature $T\lesssim 10^{4}\,{\rm K}$ and density $n_{\rm H}\lesssim 10^{-5}\,\cmjjj$, condensed cool gas of $T\sim 10^{4}\,{\rm K}$ and $n_{\rm H}\sim 10^{-2}\,\cmjjj$ in galactic halos, warm-hot ionized medium (WHIM) of $T\sim 10^{5-6}\,{\rm K}$ and $n_{\rm H}\sim 10^{-2}-10^{-4}\,\cmjjj$, and to low-density hot plasma of $T>10^{6}\,{\rm K}$ and $n_{\rm H}\lesssim 10^{-4}\,\cmjjj$, separate from the cool IGM of $T\sim 10^{4}\,{\rm K}$ and $n_{\rm H}\lesssim 10^{-5}\,\cmjjj$ which contains less than 40\% of the total mass in the diffuse gas phase \citep[{\it upper-left} panel; adapted from][for the $z=0$ Universe]{Torrey:2019}. These different phases are most effectively probed by absorption transitions associated with different ionic species, with low-ionization lines such as \ion{C}{II} and \ion{Mg}{II} tracing the cooler and higher density phase and higher-ionization lines such as \ion{O}{VI} and \ion{Ne}{VIII} tracing hotter and lower-density gas. The current mass census for different baryonic components in a Milky-Way-like halo of $M_h\approx 10^{12}\,\msun$ at $z<1$ is summarized in the {\it lower-left} panel.  The horizontal dashed line indicates the anticipated total baryonic mass from the cosmic baryon-to-dark matter ratio \citep[e.g.][]{Planck:2020}.  Colored bars represent the best-estimated mass contributions of stars \citep[e.g.,][]{Kravtsov:2018, Behroozi:2019}, the ISM \citep[e.g.,][]{Parkash:2018}, and different phases of the CGM at distinct temperature and density regimes with uncertainties indicated by the hatched regions \citep[see e.g.,][]{Faerman:2023, Qu:2023, Zhang:2024}.  Stars and the ISM combined contribute to less than 25\% of the total baryonic mass in galaxy halos, while the majority of baryons reside in the diffuse gas phases.  The schematic diagram in the {\it right} panel illustrates the current empirical understanding of the multiphase galactic atmosphere. Large-scale processes such as tidal stripping, filamentary accretion from the IGM, and thermodynamics of the hot halo interact to shape the CGM structure.  Cool, dense clumps occur primarily in the inner halo or filaments, while warm, low-density clouds occur primarily in halo outskirts, establishing a global pressure profile \citep[see also][]{Voit:2019,Qu:2023}.  These processes collectively contribute to the complex evolution of gas around galaxies, influencing star formation and galactic growth.}
  \label{fig:phases}
\end{figure}

Moving further into the distant Universe, absorption line spectroscopy has also uncovered absorption features at different wavelengths, providing signposts for different ionization states of elements like hydrogen, oxygen, silicon, and carbon (see Section \ref{sec:methods}). These ions, such as neutral hydrogen (\ion{H}{I}), \ion{C}{IV} and \ion{O}{VI}, trace gas at a broad range of temperatures.  Their distinct absorption line profiles also reflect this.  Cooler gas moves more slowly with smaller velocity dispersions, leading to a narrower line width, while hotter gas associated with more energetic outflows or more diffuse halo structures exhibits a broader line width.  

Cosmological simulations incorporating baryonic physics have consistently shown that the diffuse gas permeating the universe makes up the majority of baryonic matter.  The thermal and ionization conditions of the gas are dictated by different competing heating and cooling processes and can be broadly classified into four different regimes \citep[e.g.,][see also Fig.\ \ref{fig:phases}]{Torrey:2019}.  Specifically, in the low-density IGM, the primary heating source is photoionization by the metagalactic ionizing radiation field \citep[e.g.,][]{Madau:1995, Haardt:2012, Khaire:2019, F-G:2020}. As the Universe expands, the IGM undergoes adiabatic cooling, forming a diffuse phase with a tight correlation between gas temperature and density \citep[e.g.,][see the {\it upper-left} panel of Fig.\ \ref{fig:phases}]{Hui:1997}.  At the mean density of the present-day Universe, $n_{\rm H}\!\approx\!10^{-5}\,\cmjjj$, this unshocked gas remains at a relatively cool temperature of $T\!\sim\!10^4$ K.  At the same time, gas in overdense regions is expected to follow gravitational collapse and become shock-heated to virial temperature $T_{\rm vir}$, which scales with the mass of the bound ``halo'', $M_h$, according to $T_{\rm vir}\!\approx\!1.2\!\times\!10^6\,{\rm K}\,(M_h/10^{12}\,\msun)^{2/3}$ from the Virial theorem \citep[e.g.,][]{Mo2010}.  The gas remains hot due to inefficient cooling at low densities of $n_{\rm H}\!\lesssim\!10^{-4}\,\cmjj$.  Within the hot halos, density fluctuations induced by various dynamical processes, including IGM accretion, satellite interactions, turbulence, and feedback from active galactic nuclei (AGN) or massive stars are expected to trigger thermal instabilities with higher-density regions cooling at a faster rate \citep[e.g.,][]{Mo:1996,Maller:2004, Voit2017}.  As the temperature decreases, these regions are expected to collapse and form cooler and higher-density clumps of $T\!\sim\!10^4$ and $n_{\rm H}\!\gtrsim\!0.01\,\cmjjj$, re-establishing pressure equilibrium with the ambient hot medium.  The gas remains ionized by the metagalactic ionizing radiation field. At the interface of the cool clumps and hot medium, a transient warm-hot ionized phase of $T\!\sim\!10^{5-6}$ and $n_{\rm H}\!\sim\!10^{-3}$--$10^{-4}\,\cmjjj$ can form through shocks or turbulent mixing due to accretion or outflows interacting with the ambient medium. 

To capture the full range of physical processes that shape the properties of multiphase gas, it is essential to sample the broad dynamic range of temperatures and densities across a representative cosmological volume---a daunting challenge for both theoretical and observational studies. To the advantage of astronomers, different metal ions can serve as tracers of different gas phases, depending on the gas's chemical composition, temperature, and density \citep[e.g.,][]{OsterbrockFerland2006}. Each ion's distinctive electron configuration provides unique line transitions as empirical probes of these gas phases. For example, in cool, condensed clouds, metals are primarily singly ionized and prominent transitions include \ion{Mg}{II}, and \ion{Ca}{II}, similar to what is observed in the ISM. In contrast, more highly ionized species dominate in progressively warmer, lower-density gas and sensitive tracers include \ion{C}{IV} for warm ionized gas and \ion{O}{VI} for the warm-hot plasma \citep[Fig.\ \ref{fig:phases}; see][for a more complete list of transition]{Verner:1994a, Verner:1994b}.  Therefore, targeting a suite of ions and their associated line features provides a powerful tool for characterizing the multiphase CGM.

A census of the mass budget between different baryonic components in halos like the Milky Way is summarized in the {\it lower-left} panel of Fig.\ \ref{fig:phases}.  Stars and ISM in star-forming regions make up less than 25\% of the expected total baryonic mass from the cosmic baryon-to-dark matter ratio \citep[e.g.,][]{Planck2020}.  In contrast, the cool and warm-hot ionized phases combined in the CGM contribute to a comparable amount of mass (and likely more) found in stars.  A still larger fraction of baryonic mass is found in the hot atmosphere but with large associated uncertainties due to observational challenges.  The schematic diagram in Fig.\ \ref{fig:phases} illustrates the relative spatial distributions of these different phases in a galaxy halo, following a common pressure profile.  Condensed cool clumps occur primarily in the inner halo or along the filaments, while low-density, warm clouds are seen primarily in the halo outskirts. These clumps are embedded in a hot atmosphere stirred by various large-scale dynamical processes, including tidal stripping of satellite galaxies, IGM accretion, and outflows interacting with the hot halo. As the cool clumps move through the hot medium, turbulent mixing layers are expected to develop resulting in an additional intermediate transient phase. 

While significant progress has been made in understanding the multiphase structure of the CGM around Milky Way-like galaxies, key questions remain.  Among them are the fate of cool gas clumps and the distribution of baryons across halos of varying mass and star formation histories. Under gravity, cool, dense clumps are expected to fall through the hot halo, but whether they can survive this descent has major implications for maintaining the fuel needed to sustain star formation.  Moreover, the large uncertainties in the CGM mass budget, as shown in Fig.\ \ref{fig:phases}, suggest that more than half of the baryons may lie beyond the gravitationally bound halos. This has profound implications for the energy required to drive galactic outflows and highlights the persistent issue of the missing baryons in the universe  \citep[e.g.,][]{Shull:2012, deGraaff:2019}.


\section{Observational methods} \label{sec:methods}

To investigate diffuse multiphase gas, it is essential to consider how gas particles interact with photons. Taking hydrogen as an example, each neutral atom consists of one electron in a quantum state of energy $E_n=-13.6\,{\rm eV}/n^2$.  
To ionize a hydrogen atom and eject its electron, a high-energy photon with at least 13.6 eV (corresponding to a wavelength of $\lambda=912$ \AA\ or 91.2 nm) is required.  As noted in Section \ref{sec:phases}, the diffuse IGM is expected to be fully ionized by the metagalactic ionizing radiation field, maintaining a typical temperature of $10^4$ K 
\citep[e.g.][]{OsterbrockFerland2006}.  Under equilibrium, the photoionization and recombination maintain a stable fraction of neutral hydrogen following $n_{\rm HI}\,\Gamma=n_en_p\alpha$, where $n_{\rm HI}$, $n_e$ and $n_p$ represent respectively the neutral hydrogen, electron, and ionized hydrogen densities. The likelihood of an electron being ejected from a neutral hydrogen atom or captured by ionized hydrogen is quantified by the photoionization rate $\Gamma$ and the recombination coefficient $\alpha$, respectively. The ionization balance equation naturally leads to the expected neutral fraction, $f_{\rm HI}\equiv n_{\rm HI}/(n_{\rm HI}+n_p)$ and $n_p\approx n_e\approx n_{\rm H}$ for highly ionized gas, leading to $f_{\rm HI}\approx n_{\rm HI}/n_p=n_e\alpha/\Gamma$.  Although this neutral fraction is minuscule in the IGM ($f_{\rm HI}\lesssim 0.01$), hydrogen is the most abundant element in the Universe \citep[e.g.,][]{Payne:1925}.  These few neutral hydrogen atoms can produce observable signals, providing critical insights into the physical state of the diffuse IGM \citep[e.g.,][]{Rauch:1998}.  

\begin{figure}[t]
  \begin{minipage}[c]{0.45\textwidth}
    \includegraphics[width=\textwidth]{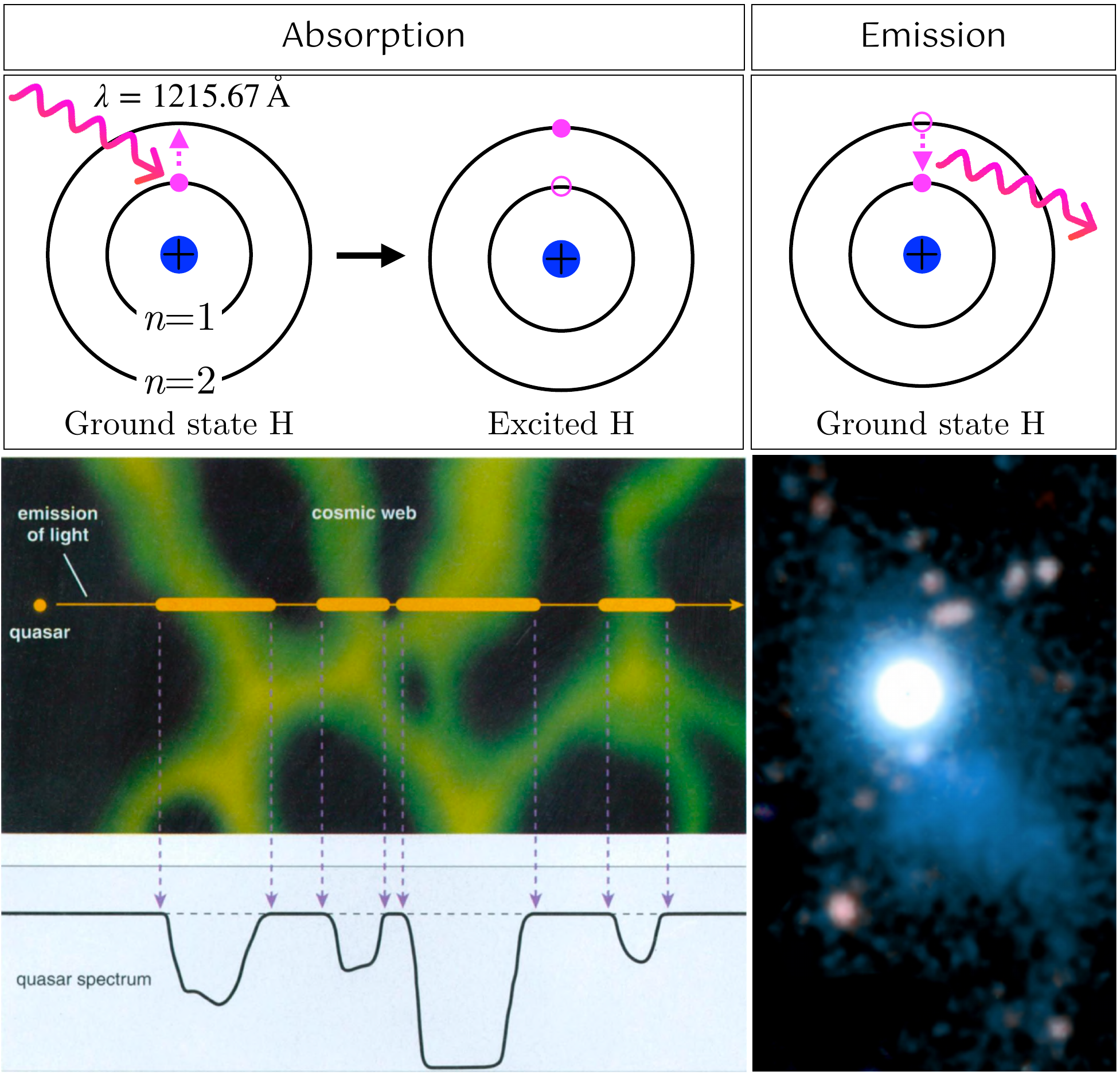}
  \end{minipage}\hfill
  \begin{minipage}[c]{0.45\textwidth}
    \caption{Mapping the diffuse cosmic gas based on absorption and emission signals.  {\it Top}: Absorption of \lya\ photons with energy matching the difference between the first and second electron levels of hydrogen atoms elevates the electron to the first excited state, attenuating the flux at \lya\ wavelength, $1215.67\, \AA$.  When the excited electron decays back to the ground state, a \lya\ photon is emitted, revealing the location of the hydrogen atom.  Gaseous streams within the cosmic filaments, made of primarily hydrogen, intercept the light from distant background quasars and are expected to attenuate the light at the \lya\ wavelength at the rest frame of the streams, resulting in absorption line features in the quasar spectra ({\it bottom left}; credit: Robert A.\ Simcoe, 2004, American Scientist; see also Fig.\ \ref{fig:qal} below).  The ``forest'' of \lya\ absorption features provides a sensitive one-dimensional probe of density fluctuations along the line of sight toward the background quasar.  Conversely, observations of \lya\ photons in emission provide two-dimensional maps of density fluctuations in the galactic atmosphere \citep[{\it lower-right}; adapted from][see also Fig.\ \ref{fig:map2d} below]{Cantalupo2014Natur}.
    } \label{fig:atomic}
  \end{minipage}
\end{figure}

A prominent spectral feature in photoionized gas is the hydrogen \lya\ line, which is produced when an electron in a neutral hydrogen atom transitions between the ground state ($n = 1$) and the first excited state ($n = 2$), releasing or absorbing energy equivalent to 10.2 eV. This energy corresponds to a wavelength of 1215.67 \AA\ (or 121.567 nm). As illustrated in Fig.\ \ref{fig:atomic}, neutral hydrogen atoms in the ground state (typical for gas cooler than $\mbox{a few} \times 10^4$ K) absorb photons at this wavelength, causing electrons to move to the $n=2$ level, resulting in the attenuation of \lya\ photon flux. Shortly after excitation, the electrons return to the ground state, emitting a \lya\ photon. Similar excitation and de-excitation processes occur for other elements, creating absorption or emission lines. The diffuse IGM and CGM emit extremely faint light when illuminated by background radiation. The interactions between gas particles and background photons vary based on density, temperature, and ionization state, producing distinct spectral features. By analyzing these features, astronomers can reconstruct 3D maps of these physical quantities, providing critical insights into the underlying cosmic structures and helping to constrain theoretical models.

\subsection{Emission measures} \label{sec:emission}

Direct imaging of faint emission from cosmic gas offers a detailed map of the cosmic matter distribution, revealing the intricate connections and interactions between this diffuse gas and galaxies where stars form. The brightness of the gas can be predicted using its emission coefficient, $j_\nu$, which quantifies the energy radiated per unit volume, per unit solid angle (measured in steradians), and per unit time. For hydrogen \lya\ emission, this is determined by the likelihood of a free electron being captured by an ionized hydrogen atom and subsequently releasing a \lya\ photon.  It can be shown following the electron cascades that for every recombination event, the probability of releasing a \lya\ photon is $\eta_\alpha=0.68$ (0.41) in optically thick (thin) cases \citep[e.g.,][]{Dijkstra2017}. The \lya\ emissivity of diffuse photoionized gas is directly related to the recombination rate of free electrons in the ionization balance equation described above, following $j_{\rm Ly\alpha}=(h\nu_{\rm Ly\alpha}/4\pi)\,n_e\,n_p\,\alpha_{\rm Ly\alpha}^{\rm eff}\,{\rm erg}\,{\rm s}^{-1}\,{\rm cm}^{-3}\,{\rm str}^{-1}$. Here $h$ is the Planck constant, $\nu_{\rm Ly\alpha}$ the \lya\ photon frequency, and $\alpha_{\rm Ly\alpha}^{\rm eff}=\eta_\alpha\,\alpha$ is the effective recombination coefficient to the $2p$ state, from where a transition to the $n=1$ is allowed \citep[e.g.,][]{Draine2011, Dijkstra2017}. For fully-ionized gas, where $n_e\approx n_p\approx n_{\rm H}$, it is straightforward to work out the anticipated surface brightness signal, integrated over a finite depth ($l_{\rm neb}$) of a nebula at redshift $z$, as ${\rm SB}_{\rm Ly\alpha}\approx 1.5\times 10^{-14}\,(l_{\rm neb}/1\,{\rm kpc})\,n_{\rm H}^2/(1+z)^4\,{\rm erg}\,{\rm s}^{-1}\,{\rm cm}^{-2}\,{\rm arcsec}^{-2}$ at $T\approx 10^4$ K, where $(1+z)^{-4}$ factor accounts for the cosmological surface brightness dimming due to expansion of the Universe.  

\begin{figure}[t]
  \centering
  \includegraphics[width=\textwidth]{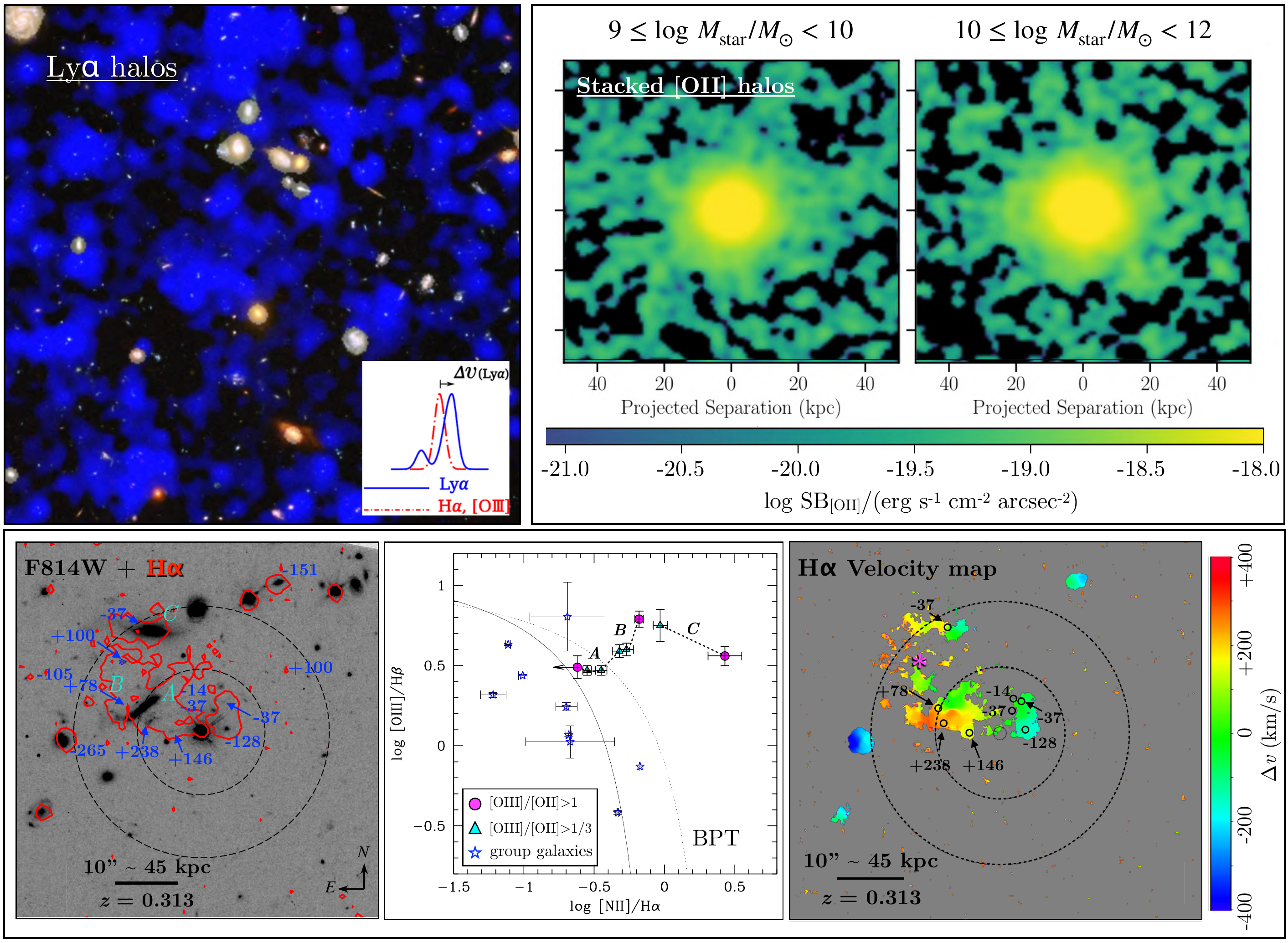}
  \caption{The spatial extent and ionization conditions of the diffuse galactic atmosphere revealed in two-dimensional line emission maps.
 Spatially extended \lya\ emitting nebulae (blue clouds in the {\it top-left} panel) are ubiquitous around high-redshift galaxies, revealing a substantial amount of gas beyond star-forming regions (reproduced from \citealt{Wisotzki2018}; see also \citealt{Cantalupo2017review, Battaia2019} for similar findings around distant quasars).   
 Due to a high cross-section with hydrogen particles, each \lya\ photon is expected to undergo a sequence of random walks both in space and frequency before escaping the parent cloud resulting in a double peak profile.  This is different from a single Gaussian line profile expected for non-resonant lines such as \ha\ or [\ion{O}{III}], resulting in an apparent offset in the observed line centroids, as illustrated in the inset (cartoon adapted from \citealt{Yang:2014}).  The two {\it top-right} panels\ (reproduced from \citealt{Dutta:2024}) summarize the current effort in expanding such observations to lower mass over an ensemble of halos.  While routine observations of individual galaxies remain challenging due to faint signals, 
 progress is being made by stacking narrow-band images targeting nebular lines over hundreds of sources.  In doing so, astronomers have detected extended line-emitting gas around galaxies with stellar mass as low as $\mstar\approx 10^8\,M_\odot$ with the detected nebular size increasing with \mstar\ at $z\lesssim 1$.  
 The {\it bottom} panels display a unique case where a giant line-emitting nebula spanning $\sim\!100$ kpc in diameter is discovered in a low-mass galaxy group with an estimated total mass of $M_h\approx 3\times 10^{12}\,M_\odot$ at $z=0.3$ \citep[adapted from][]{Chen2019}.  The observed H$\alpha$ surface brightness contours are overlaid on top of {\it Hubble Space Telescope} F814W image ({\it bottom left}).  Both the velocity field ({\it bottom right}) and line-emission morphology of the nebula support the origin of the nebula in tidal streams connecting members of the galaxy group.  This intragroup nebula is detected in multiple nebular lines, including [\ion{O}{II}], H$\beta$, [\ion{O}{III}], and [\ion{N}{II}], in addition to H$\alpha$. The observed [\ion{O}{III}]/H$\beta$ and [\ion{N}{II}]/H$\alpha$ line ratios in the BPT panel reveal an enhanced ionization in the intragroup gas (magenta circles and cyan triangles from regions $A$, $B$, and $C$ marked in the image panel) relative to the star-forming ISM (blue star symbols) in member galaxies of the group, likely due to shocks generated by the gaseous streams moving across the unseen hot intragroup medium at supersonic speeds \citep[e.g.,][]{Baldwin:1981, Kewley2019}. } 
  \label{fig:map2d}
\end{figure}

It is immediately clear that the observed strength of the emission signals depends sensitively on two factors: (1) the ionized gas density (and implicitly the intensity of the ionizing radiation field) and (2) redshift.  Because the anticipated signal depends on $n_{\rm H}^2$, the highest density peaks are expected to drive the observed signals.  It is therefore helpful to introduce a clumping factor, $C\equiv\langle n_{\rm H}^2\rangle/\langle n_{\rm H}\rangle^2$, and recast the emissivity to $j_{\rm Ly\alpha}=(h\nu_{\rm Ly\alpha}/4\pi)\,C\,\langle\,n_{\rm H}\rangle^2\,\alpha_{\rm Ly\alpha}^{\rm eff}$. Because of the $(1+z)^{-4}$ dependence, the anticipated signal fades drastically with increasing redshift.  For a 1-kpc size CGM clump of $n_{\rm H}=0.001\,\cmjjj$ \citep[e.g.,][]{Chen:2023b} at $z=2$, the expected \lya\ surface brightness signal is exceedingly faint at ${\rm SB}_{\rm Ly\alpha}\approx 2\times 10^{-22}\,{\rm erg}\,{\rm s}^{-1}\,{\rm cm}^{-2}\,{\rm arcsec}^{-2}$ for $C=1$.  Detecting this faint \lya\ emission signals from the diffuse IGM remains out of reach \citep[see e.g.,][]{HoganWeymann1987, Kollmeier2010}.  In contrast, dedicated surveys on large ground-based telescopes, ultra-long integrations ($\approx 30$-100 hours) on the 8~m Very Large Telescopes (VLT) are beginning to uncover extended \lya\ emission signals of ${\rm SB}_{\rm Ly\alpha}\approx \mbox{a few}\!\times\!10^{-20}\,{\rm erg}\,{\rm s}^{-1}\,{\rm cm}^{-2}\,{\rm arcsec}^{-2}$ out to $\approx 30$ kpc from star-forming galaxies at $z\gtrsim 3$ (Fig.\ \ref{fig:map2d}, adapted from \citealt{Wisotzki2018}; see also \citealt{Rauch2008}).  In addition, giant \lya\ nebulae of surface brightness ${\rm SB}_{\rm Ly\alpha}\gtrsim 10^{-18}\,{\rm erg}\,{\rm s}^{-1}\,{\rm cm}^{-2}\,{\rm arcsec}^{-2}$ and spatial extent $>100$ kpc are also found ubiquitous around luminous quasars \citep[e.g.,][]{Cantalupo2014Natur, Martin:2015, Borisova2016, Battaia2019}, suggesting a causal link between the presence of the cool gas ($T\sim 10^4$ K) and the quasar phase.

While optically thick gas exposed to the intense ionizing radiation from a quasar is expected to glow in \lya\ at a much higher intensity, constraining the gas properties based on the observed signals is challenging due to their uncertain origins.  Recombination radiation, resonant scattering, cooling radiation, or a combination thereof can contribute to the observed \lya\ signals.  In addition, the large absorption cross-section expected of the \lya\ photons ensures that in ionizing clouds with even a modest neutral hydrogen column density of $N_{\rm HI}\!>\!10^{13}\,\cmjj$, these photons will undergo a sequence of random walk in both space and frequency before leaving the nebula \citep[see e.g.,][for a pedagogical review]{Dijkstra2017}. \lya\ emission originating in infalling (outflowing) medium is expected to exhibit a blue-enhanced (red-enhanced) double-peak profile (see illustration reproduced from \citealt{Yang:2014} in the inset of the {\it upper-left} panel in Fig.\ \ref{fig:map2d}).  While the strong coupling between the \lya\ photons and the underlying gas motions make spectrally resolved \lya\ profiles a valuable probe of gas kinematics, constraining the gas properties requires sophisticated radiative transfer models \citep[e.g.,][]{Verhamme2006, Gronke2015}.

Non-resonant lines provide a valuable alternative, although the signals are notably weaker. Commonly seen optical nebular lines from \ion{H}{II} regions include optical recombination lines such as the hydrogen Balmer series (H$\alpha\,\lambda\,6564$, H$\beta\,\lambda\,4862$, and higher-order lines), as well as ``forbidden'' lines such as [\ion{O}{II}]\,$\lambda\lambda\,3727, 3729$, [\ion{N}{II}]\,$\lambda\lambda\,6549, 6585$, and [\ion{O}{III}]\,$\lambda\lambda\,4960, 5008$.  In particular, the hydrogen Balmer series, produced by electron transitions between the $n=2$ state and the higher orbits of neutral hydrogen atoms, also serves as a good tracer of the total baryon content.  Similar to \lya, the anticipated H$\alpha$ emissivity of ionized nebulae is $j_{\rm H\alpha}=(h\nu_{\rm H\alpha}/4\pi)\,C\,\langle\,n_{\rm H}\,\rangle^2\,\alpha_{\rm H\alpha}^{\rm eff}$ with a corresponding surface brightness of ${\rm SB}_{\rm H\alpha}\approx 2\times 10^{-15}\,(l_{\rm neb}/1\,{\rm kpc})\,C\,\langle\,n_{\rm H}\rangle^2/(1+z)^4\,{\rm erg}\,{\rm s}^{-1}\,{\rm cm}^{-2}\,{\rm arcsec}^{-2}$ for $\alpha_{\rm H\alpha}^{\rm eff}\approx 1.17\times 10^{-13}\,{\rm cm}^3\,{\rm s}^{-1}$ at $T\approx 10^4$ K \citep[e.g.,][]{Draine2011}.  Although the anticipated H$\alpha$ signals are nearly ten times fainter than \lya, these photons are not strongly coupled with the gas along the line of sight and provide a robust record of the underlying gas density and velocity.  The challenge associated with an inherently fainter signal from H$\alpha$ can be mitigated by targeting gaseous halos at lower redshifts where the effect of cosmological surface brightness dimming is significantly smaller \citep[e.g.,][]{Lokhorst2019}.  

In addition to hydrogen, metal ions in diffuse ionized nebulae are known to be crucial in regulating the thermal state of the gas by acting as effective coolants \citep[e.g.,][]{OsterbrockFerland2006, Draine2011}. These ions, due to their complex electron orbit configurations, possess several low-energy states. Collisional excitation of electrons between these states followed by radiative decay allows energy to be efficiently dissipated with the released photons. The so-called ``forbidden'' transitions occur in these metal ions when transitions between energy levels have a very low probability. As a result, the released photons from these forbidden emission lines can escape, leading to gas cooling.  Such cooling radiation expected of diffuse ionized gas provides additional tracers of gaseous halos around galaxies.

Indeed, observations of galaxies and quasars at $z\lesssim 1$, targeting these optical nebular lines have uncovered extended emission signals far beyond previously known luminous star-forming regions \citep[e.g.,][]{Epinat:2018, Rupke2019, Johnson:2024}.  Simultaneous observations of both hydrogen recombination lines and metal lines offer the additional advantage of constraining the underlying gas density, ionization condition, metallicity, and temperature based on the relative line ratios \citep[e.g.,][]{Baldwin:1981, Kewley2019}.  An example is displayed in the {\it bottom} panels of Fig.\ \ref{fig:map2d}, showcasing a giant intragroup nebula detected in multiple nebular lines, including [\ion{O}{II}], H$\beta$, [\ion{O}{III}], and [\ion{N}{II}], in addition to H$\alpha$ \citep[][]{Chen2019}.  The emission signals span $\approx\,100$ kpc in diameter, revealing gaseous streams connecting different members of the galaxy group. The detections of multiple nebular lines enable a detailed study of the ionization state of the gas using the Baldwin, Phillps, \& Terlevich (BPT) diagram \citep[][]{Baldwin:1981}.  For photo-ionized gas in star-forming regions, the line ratios are found to follow a relatively tight sequence from top-left to bottom right \citep[solid curve from][]{Kauffmann2003}, while the presence of AGN or shocks elevate the ionization state of metal ions leading to enhanced signals of ``forbidden'' lines relative to hydrogen (to the right of the dotted curve in the BPT diagram from \citealt{Kewley2001}).  In this case, a lack of AGN signatures in the group members, coupled with the emission morphology of the nebulae, supports a scenario in which the gas is shock-heated by the tidally stripped gaseous streams moving across the unseen hot intragroup medium.  This example highlights how dynamic effects due to galaxy interactions can be effective in releasing metal-enriched gas from star-forming regions in the absence of galactic superwinds \citep[see also][for examples of gas stripping in nearby galaxy clusters]{Fossati:2016}.

Expanding observations of extended nebular emissions to lower-mass halos remains a significant challenge due to the faintness of the signals. Progress is being made by stacking narrow-band images targeting nebular emission lines across hundreds of sources.  These efforts are beginning to shed light on the extended gas around galaxies with stellar mass as low as $\mstar\approx 10^8\,\msun$.  For example, enhanced \ion{Mg}{II} emission signals are detected out to 10 kpc along the polar axis in stacked images of edge-on galaxies at $z\approx 1$ \citep[e.g.,][]{Guo:2023}, revealing the role of galactic outflows in the metal transport around distant star-forming galaxies. Beyond the disks, the extent of detected nebular signals increases with increasing \mstar\ in the stacked images \citep[see the two {\it top-right} panels of Fig.\ \ref{fig:map2d}, adapted from][]{Dutta:2024}. While stacked narrow-band images offer a global view of extended nebulae around galaxies across a wide mass range, detailed studies of individual halos are essential to characterize the physical conditions of the gas and understand their relationship with the galaxies' star formation histories.  
 
\subsection{Absorption spectroscopy} \label{sec:abs}

Absorption spectroscopy complements emission surveys by offering a sensitive tool for investigating diffuse, large-scale baryonic structures in the distant Universe (e.g., \citealt{Rauch:1998,Wolfe:2005}).  As illustrated in Fig.\ \ref{fig:atomic}, photons emitted by a background source interact with the material along the line of sight, resulting in distinct absorption features in the spectrum of the background source.   Distant quasars have long been used to illuminate and study the diffuse gas content between the quasar and the observer, providing valuable insights into the low-density IGM and CGM properties. The strengths of these absorption features are dictated by the physical conditions of the gas, including gas density, temperature, ionization state, and metallicity.  By analyzing these absorption spectra, astronomers can uniformly survey the diffuse cosmic gas structure over the vast range in density and temperature predicted by simulations (see Fig.\ \ref{fig:phases}).

\begin{figure}[t]
  \centering
  \includegraphics[width=0.95\textwidth]{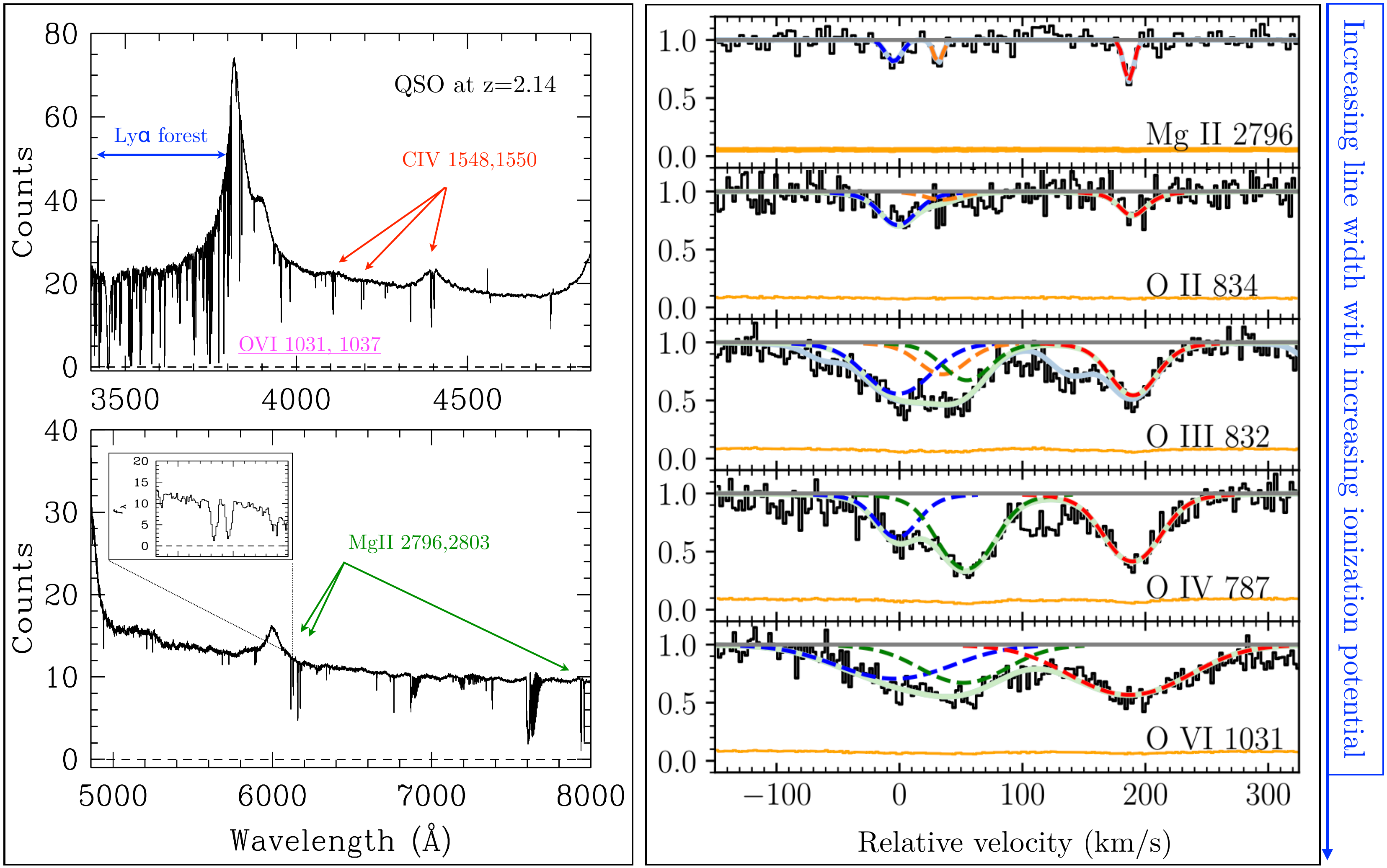}
  \caption{Quasar absorption spectroscopy provides a sensitive probe to uncover a wealth of information for diffuse multiphase gas in the galactic atmosphere, including gas density, temperature, ionization state, and kinematics, otherwise inaccessible to astronomers.  The {\it left} column displays a typical optical spectrum of a distant quasar at $z\!=\!2.14$. In addition to broad emission lines intrinsic to the quasar, such as \lya\ and \ion{N}{V} at $\approx\!3800\,\AA$, a forest of \lya\ absorption lines is observed blueward of $3800\,\AA$, revealing overdense regions spanning over ten decades in neutral hydrogen column densities $N_{\rm HI}$ at $z_{\rm abs}<z_{\rm QSO}$ along the line of sight. Many of the stronger \lya\ absorbers are accompanied with metal absorption transitions such as the \ion{O}{VI}\,$\lambda\lambda\,1031$,1037 doublet transitions, and the \ion{C}{IV}\,$\lambda\lambda\,1548$,1550 and \ion{Mg}{II}\,$\lambda\lambda\,2796$,2803 doublets (shown in the inset of the {\it bottom-left} panel). Together, the relative strengths between different ionic lines place strong constraints on the ionization state and chemical enrichment of the gas.  In addition, high-resolution absorption spectra provide additional power to resolve multiphase gas based on the observed line profiles. In particular, the absorption line widths of individual ions, typically described by the Doppler parameter $b_I$ to quantify the line-of-sight velocity dispersion, are often found to increase with the ionization potential (I.P.) of the ion. An example is displayed in the {\it right} panels, which demonstrate the progression of broader linewidths of oxygen ions at higher ionization stages \citep[adapted from][]{Cooper:2021}. The correlation between line width and ionization potential remains when incorporating all available lines from hydrogen, carbon, and oxygen through iron \citep[e.g.,][]{Qu:2022}.  
  Because different particles at different ionization stages are expected to occur in different temperature and density phases (see Fig.\ \ref{fig:phases}), the observed increasing line width with increasing ionization stages provides an independent confirmation for the presence of multiphase gas in the galactic atmosphere \citep[see also][]{Liang:2016,Kumar:2024}. }
  \label{fig:qal}
\end{figure}

An example of a typical optical spectrum of distant quasars is shown in Figure~\ref{fig:qal}. At redshift $z_{\rm QSO}=2.14$, this quasar displays prominent broad emission lines from \lya, \ion{Si}{IV}, \ion{C}{IV}, and \ion{C}{III} \citep[e.g.,][]{VandenBerk:2001}, all redshifted to the optical spectral window.  Notably, the \lya\ forest appears blueward of the \lya\ emission line at 4800 \AA, consisting of numerous \lya\,$\lambda$\,1215 absorption lines of varying strengths caused by intervening overdense regions along the QSO sightline at $z_{\rm abs}\lesssim z_{\rm QSO }$ (see also Fig.\ \ref{fig:atomic}).  The degree of flux attenuation in each absorption line quantifies the total gas mass in these intervening regions, following the relation $-\ln\,(f_{\rm obs}/f_{\rm intrinsic})=N_{\rm HI}\,\sigma_{\rm Ly\alpha}$. Here $N_{\rm HI}$ is the \ion{H}{I} column density (the number of neutral hydrogen atoms per unit area) and $\sigma_{\rm Ly\alpha}=3.3\times 10^{-14}\,{\rm cm}^2$ is the absorption cross-section of \lya\ photons at $T\approx 10^4$ K.  For a cloud of density $n_{\rm H}\approx 0.01\,\cmjj$ and size 100 pc, typical of the cool gas seen in the Milky Way Halo \citep[e.g.,][]{Putman:2012}, a neutral fraction of 1\% would lead to $N_{\rm HI}\approx 3\times 10^{16}\,\cmjj$. The large absorption cross-section indicates that even a small fraction of neutral hydrogen particles in these clouds would lead to saturated \lya\ absorption lines.  At the same time, the \lya\ forest spectrum reveals a vast range of $N_{\rm HI}$, from $N_{\rm HI}\!\gtrsim\!10^{20}\,\cmjj$ resembling neutral interstellar matter, 
to 
$N_{\rm HI}\lesssim 10^{13}\,\cmjj$ in the IGM, with the majority the lines showing $N_{\rm HI}\lesssim 10^{16}\,\cmjj$ \citep[e.g.,][]{Rauch:1998}.  The $N_{\rm HI}$ distribution function of the \lya\ forest \citep[e.g.,][]{Ribaudo:2011} provides the clearest indication yet for a highly ionized IGM and CGM (e.g., \citealt{Gunn:1965}, see also \citealt{Rauch:2000} for a review).

A significant fraction of \lya\ absorbers with $N_{\rm HI}>10^{14}\,\cmjj$ also exhibit associated metal absorption lines, indicating that the gas has been enriched with metals \citep[e.g.,][]{Cowie:1995AJ, Schaye:2000}.  The most prominent metal absorption features include the \ion{O}{VI}\,$\lambda\lambda$\,1031, 1037 doublet transitions within the \lya\ forest, as well as the \ion{C}{IV}\,$\lambda\lambda$\,1548, 1550 and \ion{Mg}{II}\,$\lambda\lambda$\,2796, 2803 doublets.  Additionally, there is a series of low-ionization transitions such as \ion{C}{II}, \ion{Si}{II}, and \ion{Fe}{II} \citep[e.g.,][]{Becker:2015, Dey:2015}.  These doublet transitions arise from lithium-like ions with two closely spaced fine-structure levels above the ground state. Their fixed doublet ratios, determined by the transition probabilities between the ground state and these fine-structure levels, along with precise transition wavelengths, significantly enhance the reliability of their identification (see the inset in the {\it lower-right} panel of Fig.\ \ref{fig:qal}).  

The presence of metals in the diffuse CGM and IGM presents one of the greatest puzzles in modern astrophysics because elements heavier than lithium are primarily formed in stellar interiors \citep[e.g.,][]{Nomoto:2013}. Transporting these heavy elements from star-forming regions on parsec scales into the gaseous halos on scales of 100 kpc and beyond over the final lifetime of the Universe would require ultra-high-speed winds from stars and supermassive black holes (SMBHs).  While supergalactic winds are seen in some nearby galaxies \citep[e.g.,][]{Veilleux:2005}, the extent of the signals reaches $\approx 10$-20 kpc, far smaller than the cosmological distance required to explain the amount of metals seen beyond star-forming regions \citep[e.g.,][see also Fig.\ \ref{fig:cgmim}]{Scannapieco:2002, Peeples:2014}.  The typical size of metal-enriched halos, $r_Z$, necessary to explain the observed number density, $n(z)$, of metal absorbers per unit redshift interval per line of sight, can be estimated following a halo cross-section weighted space density of galaxies,
$n(z) = (c/H_0)\,E(z)\,\int_0^\infty
\Phi(L,z)\,\pi\,r_Z^2(L)\,\kappa\,dL\,$,
where $c$ is the speed of light, $H_0$ is the Hubble constant, representing the current expansion rate of the Universe, $L$ is the intrinsic luminosity of galaxies in the unit of the characteristic luminosity $L_*$, $\Phi(L,z)$ is the galaxy luminosity function describing the number density of galaxies per unit comoving volume per luminosity interval, $\kappa$ is the covering fraction of metal-enriched gas in the halo, and $E(z)\equiv(1 + z)^2/\sqrt{\Omega_{\rm M}\,(1+z)^3 + \Omega_{\rm \Lambda}}$ accounts for the cosmological expansion of the cosmos. 
Here, it is assumed that more luminous galaxies are on average more massive and possess larger chemically enrichment halos \citep[e.g.,][]{Tinker:2008}, which is supported by empirical observations \citep[e.g.,][]{Chen:2010, Churchill:2013}.  Owing to the success of numerous deep galaxy surveys, the galaxy luminosity functions are well determined over a broad redshift range and star formation histories \citep[e.g.,][]{Behroozi:2019}, and the expected size of metal-enriched halo from the observed number density of metal absorbers along random quasar sightlines exceeds 100 kpc for $\kappa=50$\% \citep[e.g.,][]{Chen:2017}.

Irrespective of the physical processes that drive the chemical enrichment in underdense regions, the presence of metal absorption lines in the \lya\ forest provides valuable diagnostics for the underlying physical and thermodynamic conditions of the gas \citep[e.g.,][]{Verner:1994a, Rauch:1997, Gnat:2017}.  As noted in Section \ref{sec:phases} and Fig.\ \ref{fig:phases}, each heavy element offers multiple ionization stages, and higher ions are expected to be more dominant in warmer and lower-density CGM.  The ionization state of a photoionized gas is commonly described by a single ionization parameter, $U$, which represents the number of incident ionizing photons per hydrogen atom. For a given radiation field characterized by a total flux of hydrogen-ionizing photons, $\phi$, with energy $\geq 13.6$ eV, the $U$ parameter is inversely proportional to the hydrogen number density, $n_{\rm H}$, following the relation $U\equiv \phi/c\,n_{\rm H}$.  A higher $U$ corresponds to a more highly ionized gas and an increased abundance of highly ionized species. Therefore, the observed ion ratios place constraints on $U$, and the gas density, $n_{\rm H}$, can be inferred by assuming a fiducial ionizing spectrum \citep[e.g.,][]{Haardt:2012, Khaire:2019, F-G:2020}. Consequently, the relative abundances of different ions provide direct constraints on the gas density and ionization conditions \citep[see e.g.,][]{Chen:2017metal}. 

In addition, the observed absorption line width, traditionally described by the Doppler parameter, $b$, directly measures the velocity distribution of particles within the absorbing cloud along the line of sight.  The Doppler velocity width measures the particles' average relative line-of-sight velocity offset and is related to the velocity dispersion following $b=\sqrt{2}\,\sigma_\varv$.  For pure thermal motions, the velocity dispersion reflects the gas temperature and the thermal $b_{\rm th}$ value is expected to be $b_{\rm th}=\sqrt{2k_BT/m_{\rm I}}$, where $k_B$ is the Boltzmann constant and $m_{\rm I}$ is the mass of the ions.  Therefore, it follows that under pure thermal broadening, absorption features originating in warmer gas would appear broader, while transitions from heavier ions would appear narrower.
At the same time, the diffuse absorbing clouds in the CGM and IGM are part of a dynamic system characterized by a mixture of coherent flows and chaotic turbulence. These bulk motions can contribute significantly to the observed line widths in absorption spectra. All particles, regardless of their mass, are influenced by these motions, resulting in a constant non-thermal component, $b_{\rm nt}$, in the observed line width.  Therefore, the total observed line width, $b_{\rm obs}$, is the result of both thermal and non-thermal broadening effects and can be expressed as $b_{\rm obs}=\sqrt{b_{\rm th}^2+b_{\rm nt}^2}=\sqrt{2k_{\rm B}T/m_{\rm I}+b_{\rm nt}^2}$.  Comparing the observed line widths of two ions with significantly different masses enables astronomers to disentangle the contributions of thermal and bulk flows, thereby providing robust constraints on the gas temperature and the magnitude of non-thermal motions within the absorbing structure \citep[e.g.,][]{Rauch:1996, Rudie2019}. 

Observations have frequently shown that heavier ions produce narrower absorption line widths compared to lighter ions. For instance, Fig.\ \ref{fig:qal} highlights a narrower \ion{Mg}{II} absorption line relative to the corresponding \ion{O}{II} feature (top two right panels). Additionally, the line widths of oxygen ions increase with ionization potential, $E_i$.  Generally, ions with ionization potentials $E_i > 20$ eV show on average broader line widths ($b_{\rm obs}\gtrsim 20\,\kms$), while lower ionization states tend to exhibit narrower line widths ($b_{\rm obs}\lesssim 10\,\kms$; e.g., \citealt{Cooper:2021, Qu:2022}). This correlation between line width and ionization potential aligns with expectations of thermal broadening, as higher-ionization species trace warmer, lower-density gas (e.g., Fig.\ \ref{fig:phases}). 

By analyzing both the relative absorption strengths and velocity widths of ions across different ionization states and different masses, strong constraints can be placed on the physical and thermal properties of the absorbing gas. Therefore, absorption spectroscopy provides a powerful tool for characterizing the underlying structure and dynamics of diffuse multiphase gas.
 
\section{Known empirical properties of the multiphase CGM} \label{sec:empirical}

Quasar absorption spectroscopy has long been a powerful tool for obtaining detailed insights into the diffuse, multiphase CGM \citep[e.g.,][for reviews]{Tumlinson2017, Chen2017}. These studies typically utilize background quasars to probe the gas associated with galaxies along the line of sight, with absorption features revealing the properties of their extended atmosphere.  However, due to the rarity of luminous quasars---approximately one bright quasar ($g\!\lesssim\! 18$ mag) per square degree (e.g., \citealt{Richards2006})---these investigations have been largely restricted to a single probe per galaxy, particularly for galaxies beyond the local universe \citep[see][for studies of the Milky Way, M31, and Local Group dwarfs]{Richter:2017, Lehner:2020, Zheng:2020}.  While multiply-lensed quasars offer an exciting avenue for expanding these studies from one-dimensional to two-dimensional mapping, the number of available systems remains limited (e.g., \citealt{Rauch2001, Chen2014, Zahedy2016, Rubin2018}; see also \citealt{Lopez2018} for an example of using a highly-magnified distant galaxy as multiple probes). Nonetheless, even with one quasar per halo, statistical studies employing a large sample of quasar-galaxy pairs allow astronomers to establish a global understanding of the diffuse CGM based on ensemble averages.

\subsection{Global distributions} \label{sec:kappa}

\begin{figure}[t]
  \centering
  \includegraphics[width=0.9\textwidth]{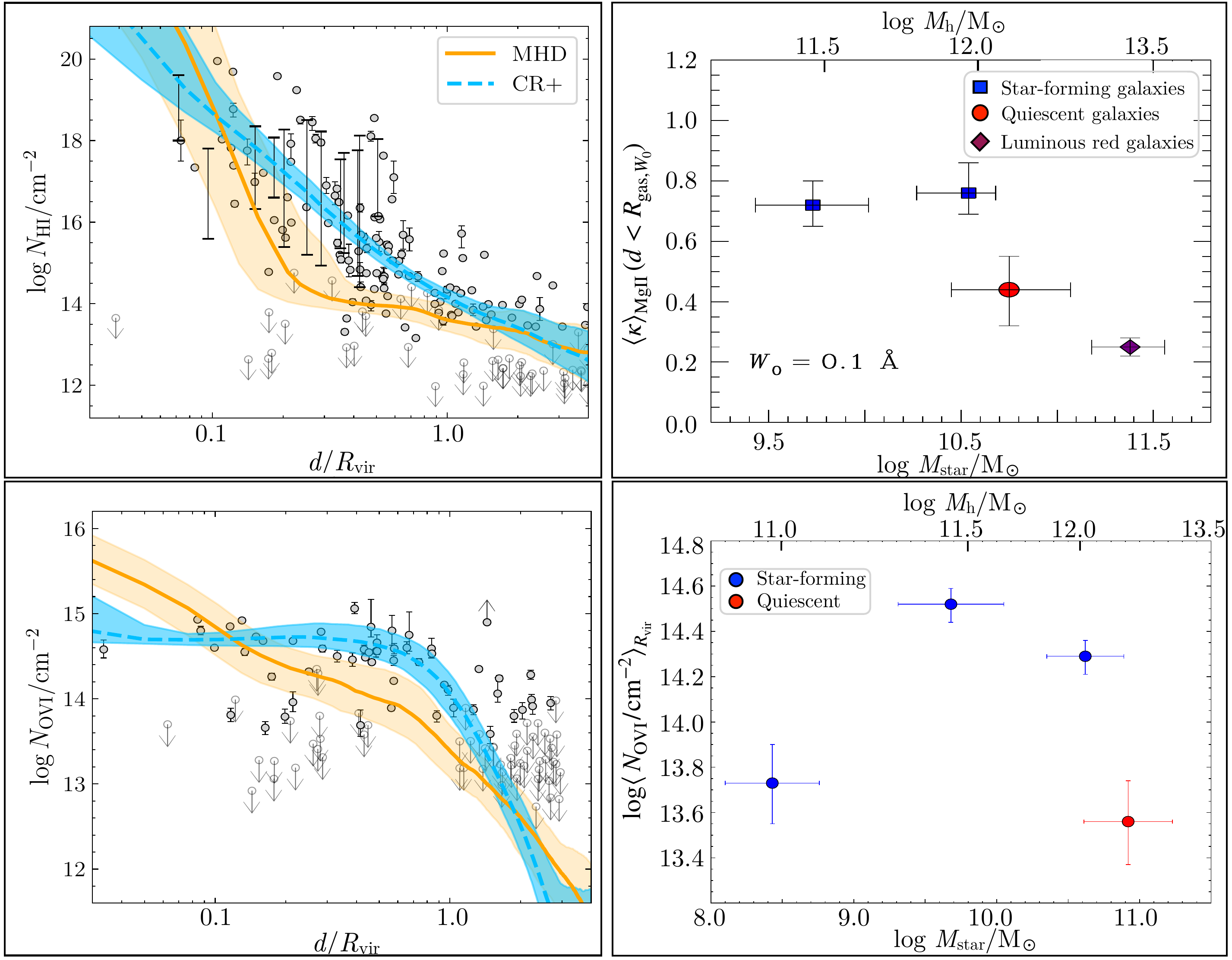}
    \caption{Empirical knowledge of the spatial distribution of multiphase gas in the galactic atmosphere from ensemble studies utilizing absorption spectroscopy. Using a sample of close galaxy and quasar pairs, strong constraints on the spatial properties of gaseous halos can be established based on the absorption strength imprinted in the spectrum of the background quasars. The data points in the {\it left} panels are compiled from a collection of literature samples for galaxies with stellar mass exceeding $\log\,\mstar/\msun=9.5$ \citep[e.g.,][]{Qu:2024}.  The column densities of low ionization state species such as \ion{H}{I} steadily decline with increasing projected distance from galaxies.  In contrast, highly ionized species such as \ion{O}{VI} display a flatter projected column density profile.  The blue and orange curves are predictions from magnetohydrodynamic (MHD) simulations and simulations incorporating cosmic ray (CR) feedback, with the bands representing the 68\% interval around the median values \citep[adapted from][]{Ji:2020}. In addition to the radial profiles of different atomic species in gaseous halos, an important observation is a non-unity covering fraction of these species, as indicated by the non-detections with downward arrows in the {\it left} panels.  The {\it upper-right} panel displays the mean covering fraction of \ion{Mg}{II} absorber with rest-frame absorption equivalent width greater than $W_0=0.1\,\AA$ within a fiducial gaseous $R_{\rm gas}$.  A general declining trend is seen with increasing mass and reduced star formation activities over the galaxy population \citep[e.g.][]{Huang:2021}. 
    A reduced covering fraction would translate directly to a reduced mean absorption column density.  This is seen in the {\it lower-right} panel for the observed mean \ion{O}{VI} column density, which declines from $\log\,\langle N({\rm OVI})/\cmjj\rangle\approx 14.5$ in halos of $\log\,\mstar/\msun\approx 9.7$ to $\log\,\langle N({\rm OVI})/\cmjj\rangle\approx 13.6$ in halos of $\log\,\mstar/\msun\approx 11$ (e.g., \citealt{Qu:2024}; see also \citealt{Zahedy:2019}, \citealt{Tchernyshyov:2022}).   The observed partial covering of different species as a function of galaxy mass and star formation history provides independent constraints for the baryon cycle in the galactic atmosphere.}
  \label{fig:kappa}
\end{figure}

As a first step, strong constraints on the spatial properties of the galactic atmosphere can be derived using a sample of galaxy and quasar pairs that span a range in projected separations ($d$).  By analyzing the absorption features imprinted in the spectra of background quasars, observers can examine the incidence and total absorption strengths of different species as a function of distance from the associated galaxies (Fig.\ \ref{fig:kappa}).  Despite being an ensemble average over many galaxies, these spatial profiles serve as a baseline characterization for the changing ionization condition and chemical enrichment in the galactic atmosphere.  Investigating how these profiles vary with the galaxy mass, star formation history, and environments provides key insights into the origin and primary drivers that dictate the physical state of the gas.

For example, the observed integrated column density profiles as a function of projected distance for different atomic species offer the first look at the extent of gas around galaxies through different tracer particles.  The {\it left} panels of Fig.\ \ref{fig:kappa} display the current empirical understanding of the extended \ion{H}{I} and \ion{O}{VI} gaseous envelopes around galaxies of $\log\,\mstar/\msun\gtrsim 9.5$.  As described in Section \ref{sec:methods}, the sizes of the gaseous atmosphere are expected to scale with halo mass \citep[see e.g.,][]{Chen:2010, Churchill:2013}, normalizing the projected distance by the virial radius of the halo $R_{\rm vir}$ removes this intrinsic mass scaling and places all galaxies of different masses on the same scale.  Low ionization state species such as \ion{H}{I} display a more rapidly declining profile with increasing distance than more highly-ionized species such as \ion{O}{VI} from galaxies.  The data points included in the panels are based on a collection of literature samples \citep[see][for a detailed description]{Qu:2024}.  Observations of additional low ions around galaxies of a wide range of star formation histories, such as \ion{Mg}{II}, \ion{C}{II}, and \ion{Si}{II}, can be found in \cite{Chen:2010}, \cite{Werk2014}, \cite{Liang:2014}, and \cite{Huang:2021}.   

The apparent difference in the projected column density profiles of \ion{H}{I} and \ion{O}{VI} reveals that these two species do not generally occupy the same regions within a galaxy halo. This discrepancy underscores the importance of targeting a suite of ions to develop a comprehensive and robust characterization of gaseous halos. Each ion traces gas in different ionization states and densities, meaning that no single ion can capture the full complexity of the halo. In addition, the observed differences between ions in various ionization stages point to varying ionization conditions across the halo. These spatial ionization profiles serve as key observables, helping to distinguish between competing theoretical models. 
Comparisons between simulations with and without cosmic ray (CR) feedback have indeed shown that models incorporating CR feedback better capture the large-scale distributions of both \ion{H}{I} and \ion{O}{VI} around star-forming galaxies (\citealt{Ji:2020}; blue versus orange curves in Fig.\ \ref{fig:kappa}).  While the success suggests cosmic ray feedback may be a dominant driver in regulating the diffuse halo gas, it is not the only successful model.  Other mechanisms, including AGN feedback, have also been shown to explain the observed column density profiles well \citep[e.g.,][]{Oppenheimer:2016, Nelson:2018}).  Additional observables are therefore necessary to improve the discriminating power.

In addition to comparing the spatial profiles of different species, a clear feature in the observations that has not been accounted for is the presence of non-detections at $d<R_{\rm vir}$.  These sightlines pass through regions with particularly low densities of absorbing materials in galaxy halos, indicating a patchy distribution of these materials, particularly metal-enriched gas. 
Therefore, investigating how the mean covering fraction of metal-enriched gas changes with halo mass provides further insight into how halo gas properties are coupled with halo growth.  Observations show that the covering fraction of chemically-enriched, photoionized gas traced by \ion{Mg}{II} varies with both galaxy mass and star formation history, from $\langle\kappa\rangle_{\rm MgII}\approx 80$\% in Milky-Way like star-forming halos to $\langle\kappa\rangle_{\rm MgII}\approx 20$\% in massive halos hosting luminous red galaxies ({\it top-right} panel of Fig.\ \ref{fig:kappa}).  These luminous red galaxies (LRGs) resemble nearby elliptical galaxies with little ongoing star formation \citep[e.g.,][]{Eisenstein:2001} and represent a homogeneous sample of massive quiescent systems.

Examining the mean absorption properties of highly ionized species provides additional clues.  This is shown in the {\it lower-right} panel of Fig.\ \ref{fig:kappa}, which presents the mean \ion{O}{VI} column density, $\langle N({\rm OVI})\rangle_{\rm R_{\rm vir}}$, for various mass bins.  The mean \ion{O}{VI} column density per mass bin is an integral of the product of $N({\rm OVI})$ and covering fraction across the projected area within the halo radius, $\langle N({\rm OVI})\rangle_{\rm R_{\rm vir}} = \int_0^{R_{\rm vir}}\,N({\rm OVI},d)\,\kappa_{\rm OVI}(d)\,2d\,\delta d/R_{\rm vir}^2$.  For a flat column density profile that reaches a maximum of $N_0$ ({\it lower-left} panel of Fig.\ \ref{fig:kappa}), the observed mean column density translates to a mean gas covering fraction following $\langle N({\rm OVI})\rangle_{\rm R_{\rm vir}}=N_0\,\langle\kappa\rangle_{{\rm OVI}, {\rm R_{\rm vir}}}$.  Surveys of \ion{O}{VI}-bearing CGM have uncovered $\log\,N_0/\cmjj\approx 14.7$ for Milky-Way like star-forming halos, while $\log\,N_0/\cmjj\approx 14.0$ for massive quiescent galaxies \citep[see][]{Tumlinson:2011, Qu:2024}.  At the same time, $\log\,\langle N({\rm OVI})/\cmjj\rangle_{\rm R_{\rm vir}}$ is found to be $\approx 14.5$ in halos of $\mstar\approx 10^{10}\,\msun$ and $\approx 13.6$ in halos of $\mstar\approx 10^{11}\,\msun$, indicating a rapidly declining $\langle\kappa\rangle_{{\rm OVI}{\rm R_{\rm vir}}}$ with increasing mass. Different from \ion{Mg}{II}-bearing gas, however, the mean \ion{O}{VI} column density also shows a notable decline to $\langle\,\log\,N_{\rm OVI}/\cmjj\rangle\approx 13.8$ around low-mass dwarf galaxies \citep[e.g.,][]{Qu:2024}.
The observed decline in both $\langle\kappa\rangle_{\rm MgII}$ and $\langle\kappa\rangle_{\rm OVI}$ can be attributed to either a lack of starburst outflows in massive quiescent galaxies, a higher destruction rate of CGM clumps in massive halos, or a combination of both factors. 

A key distinguishing feature of starburst-driven outflows is their non-spherical distribution in the halos of galaxies with well-formed, star-forming disks \citep[e.g.,][]{Veilleux:2005}.  Galactic-scale outflows are expected to travel preferentially along the polar axis where the gas experiences the least resistance, while accretion of the IGM is likely to proceed along the disk plane \citep[e.g.,][]{Shen:2013, Schneider:2020}.  The expectation that metal absorption strengths would be enhanced along the polar axis of galaxies, if outflows predominantly drive metals, is complicated by the realities of the CGM.  For example, the M81 group (Fig.\ \ref{fig:cgmim}) presents a case where stripped gas, inflows, and outflows mix in a way that doesn't align neatly with the major-minor axis distinction. Instead, the chaotic distribution of diffuse gas emission signals suggests that the interplay of multiple processes complicates our ability to isolate outflows based solely on the axis orientation.  Observational studies on CGM absorption in disk galaxies have also yielded mixed results. While some observations show enhanced metal absorption along the polar axes (e.g., \citealt{Bordoloi:2011}, and recall such anisotropic distribution of metal-enriched gas detected in emission from \citealt{Guo:2023}), consistent with outflow-driven models, others have found more isotropic distributions or stronger absorption along the galaxy's plane \citep[e.g.,][]{Martin:2019}. This discrepancy may result from differences in galactic environments and/or measurement uncertainties. To establish a comprehensive understanding of the diffuse galactic atmosphere, it is necessary to go beyond a simple cross-section analysis and spatially and spectrally resolve detailed physical properties of the multiphase gas.  While this is still challenging with current facilities (see Section \ref{sec:future}), significant progress has been made with spectrally resolved absorption-line studies.


\subsection{Resolved density, size, and metallicity structures} \label{sec:resolve}

In recent years, high-resolution absorption spectroscopy of the diffuse CGM has enabled astronomers to resolve small-scale features in galaxy halos and examine how the complex multiphase gas physics on small scales is influencing galaxy growth.  Recall that the observed line width is characterized by the Doppler velocity width, $b_{\rm obs}=\sqrt{2}\,\sigma_\varv$ and both thermal and non-thermal motions contribute to the observed line broadening.  For a fiducial gas temperature of $T\sim 10^4$ K, the expected thermal line width is $b_{\rm th}=\sqrt{2k_BT/m_{\rm I}}\approx 12.9$ \kms\ for hydrogen, which translates to a required spectral resolution in velocity of full-width-at-half-maximum (${\rm FWHM}_\varv)\,\approx 20$ \kms\ to resolve individual hydrogen components and ${\rm FWHM}_\varv\approx 6$ \kms\ for heavier ions in the absence of bulk motions.   The recent progress in resolved component studies is enabled by combining optical echelle spectroscopy on large ground-based telescopes with high-resolution far-ultraviolet spectra obtained using the {\it Hubble Space Telescope} to cover a large number of elements in different ionization stages expected in the diffuse multiphase gas (e.g., Fig.\ \ref{fig:phases}; see also \citealt{Verner:1994a}) and to provide the maximal dynamic range in particle mass for decomposing the multiphase components. 

An example is shown in Fig.\ \ref{fig:qal}, where an absorption system is resolved into four distinct components within a line-of-sight velocity interval of $\approx 200$ \kms\ (highlighted in blue, orange, green, and red from 0 to $\approx +200$ \kms\ in relative velocities).  While all four components exhibit progressively broader velocity widths with increasing ionization stages of oxygen ions as noted in Section \ref{sec:abs}, the corresponding \ion{Mg}{II} components are notably narrower than \ion{O}{II}, a lighter element sharing a similar ionization potential, indicating that the observed line widths of these low-ionization species are driven by thermal motions with a relatively cool phase of $T\approx 10^4$ K.  The broader widths observed in \ion{O}{III} and \ion{O}{IV} suggest the presence of a warmer phase.   Indeed a spectrally-resolved photoionization analysis showed that these intermediate ions likely originate in low densities of $\log\,n_{\rm H}/\cmjjj\approx -3.9$, while the lower ions trace higher-density regions of $\log\,n_{\rm H}/\cmjjj\approx -2.3$ \citep[][]{Cooper:2021}. The much broader line widths observed in \ion{O}{VI} indicate the presence of a third phase at a significantly higher temperature of $T\gtrsim 10^6$ K in at least three of the four components.  This example illustrates how a careful analysis of the absorption profiles of different ions allows astronomers to deblend and decompose multiphase gas, although the connections between these phases remain uncertain.  While the matching velocity centroids between the cool and warm phases observed in the blue and red components suggest that they share the same density structures with the warm phase arising in the interface between the dense core and the ambient hot halo, it is also possible that the warm phase arises in a physically distinct region, likely in the halo outskirts, along the line of sight (see the cartoon illustration in Fig.\ \ref{fig:phases}).


\begin{figure}[t]
  \centering
  \includegraphics[width=\textwidth]{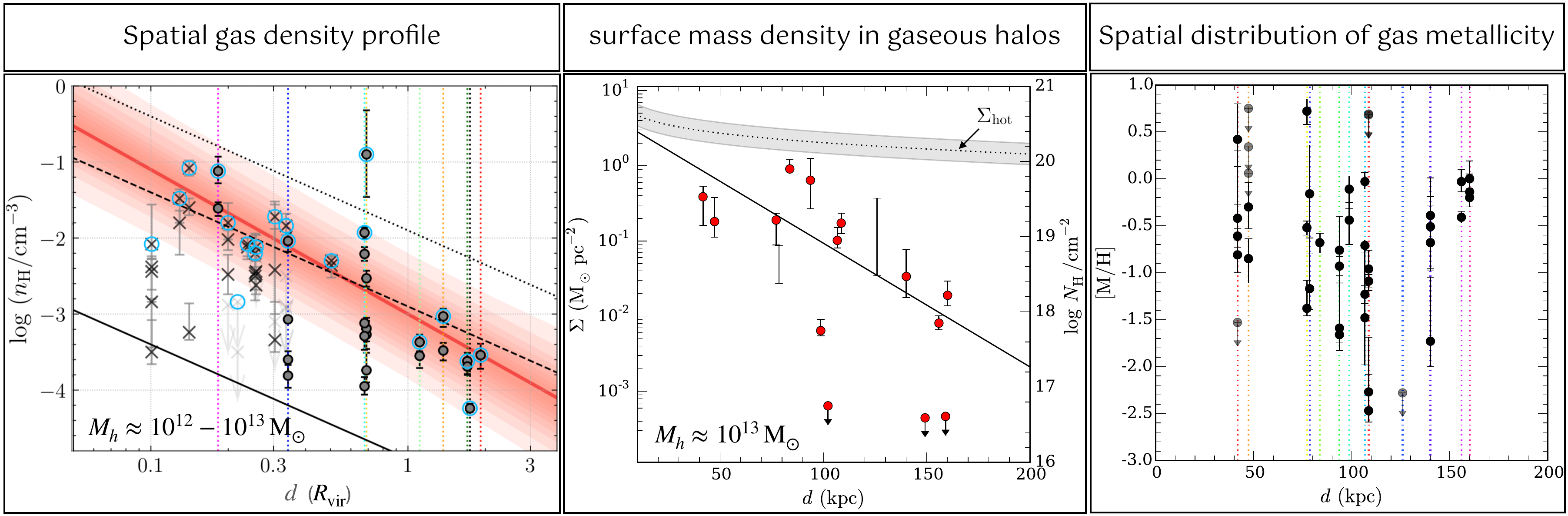}
  \caption{Global density (pressure), mass, and chemical enrichment profiles established for the diffuse galaxy halos at $z\lesssim1$ from absorption-line observations. The {\it left} panel shows the density fluctuations resolved in absorption spectroscopy for the cool gas of temperature $T\sim 10^4$ K along individual sightlines (data points connected by a vertical, dotted line) as a function of projected distance in units of halo radius $R_{\rm vir}$ (adapted from \citealt{Qu:2023}; see also \citealt{Zahedy:2019}). The large dispersion in gas density between resolved clumps can be understood by the projection effect of different clumps at different radii intercepted by each quasar sightline.  Such an explanation is supported by the gradual decline of the upper envelope of the density distribution (red curves), with quasar sightlines at large $d$ probing the low-density outskirts of galaxy halos.  It is also consistent with the density profile inferred for the hot phase \citep[black solid line; e.g.,][]{Li:2017,Singh:2018}. With a typical temperature of $T\sim 10^4$ expected for the cool photoionized gas and $T\sim 10^6$ K for the hot medium, the observed factor of $\sim 100$ difference in the global density profiles between the two phases suggests that the galactic atmosphere is in a quasi-equilibrium pressure balance, following the ideal gas law $P=n_{\rm H}k_BT$ (see e.g., Fig.\ \ref{fig:phases}).  Independent of the adopted scenario to explain the observed scatter in cool gas density, the observed density distributions over an ensemble of halos translate directly to the mean gas surface mass density in the cool phase.  The {\it middle} panel shows a comparison between the inferred surface mass density profiles of the cool (red circles and solid line) and hot ($T\gtrsim10^6$\,K; dotted curve) CGM of high-mass quiescent galaxies from \cite{Zahedy:2019}, showing a diminishing presence of cool gas toward the outskirts of these massive halos. For massive galaxies residing in halos with $M_h\approx10^{13}\,\msun$, the vast majority ($\gtrsim90\%$) of the mass of the CGM resides in the hot phase.  Similar to gas density, the inferred gas metallicity in the cool CGM also exhibits a large scatter ({\it right} panel, reproduced from \citealt{Zahedy:2019}; see also \citealt{Zahedy2021, Cooper:2021, Kumar:2024}). Measurements of gas clouds identified along a single sightline are connected by a vertical dotted line to guide the visual inspection. The wide range of metallicities seen in the CGM of individual galaxies underscores the complex chemical enrichment history in individual galaxy halos.}
  \label{fig:global}
\end{figure}

Spectrally-resolved photoionization analyses of the multiphase CGM have uncovered remarkable insights into its complex structure \citep[e.g.][]{Zahedy:2019, Zahedy2021, Cooper:2021, Kumar:2024}.  A common feature is the detection of multiple clumps along each sightline with a characteristic density of $n_{\rm H}\sim 10^{-2}\,\cmjjj$, while a large spread of nearly two orders of magnitude between clumps is also observed ({\it left} panel of Fig.\ \ref{fig:global}).  Such a large dispersion in gas density can be understood as the result of the projection effect, where quasar sightlines pass through clumps located at different distances from the galaxy center \citep[e.g.,][]{Zahedy:2019, Voit:2019, Qu:2023}.  The general trend shows that closer sightlines encounter gas across a broad range of densities, while those probing the outer CGM uncover mostly lower-density gas. This physical picture is corroborated by the observed general decline in the upper envelope of the density profile.  Moreover, this explanation is also consistent with the inferred density profile of hot CGM \citep[black solid line; e.g.,][]{Li:2017,Singh:2018}. With typical temperatures of $T\sim 10^4$ K for the cool CGM gas (see Section \ref{sec:phases}) and $T\sim 10^6$ K expected for the hot medium, the observed factor of $\sim 100$ difference in the global density profiles between the two phases suggests that the CGM is in roughly in thermal pressure equilibrium, following $n_{\rm H}^{\rm cool}T_{\rm cool}\approx n_{\rm H}^{\rm hot}T_{\rm hot}$.  However, the data are sufficiently noisy that the presence of non-pressure support in the multiphase gas from the magnetic field and cosmic rays cannot be ruled out \citep[e.g.,][]{Ji:2020, Ruszkowski:2023}, particularly for the intermediate phase with a possible origin in the boundary laters.  

The inferred density distributions directly translate to the mean gas surface mass density of the cool CGM, as shown in the {\it middle} panel of Fig.\ \ref{fig:global} for a sample of massive quiescent galaxies \citep[][]{Zahedy:2019}. For these high-mass galaxies, the radial profile of surface mass densities exhibits a strongly declining presence of cool gas toward larger radii in the halo.  In contrast, the surface mass densities of the hot halo inferred from a joint analysis of X-ray and Sunyaev-Zel'dovich signals \citep[e.g.,][]{Sunyaev:1980} appear to be significantly flatter (\citealt{Singh:2018}; light grey band in the {\it middle} panel of Fig.\ \ref{fig:global}), suggesting that the hot halo is more spatially extended than the cool gas.  Integrating the surface mass density profile over the projected area directly constrains the gas mass in the cold and hot phases.  Within $d\approx 160$ kpc, it was found that the cool phase contains a total mass of $M_{\rm cool}\sim 1.5\times 10^{10}\,\msun$, which is about 10\% of the total mass in the hot phase \citep[][]{Zahedy:2019}. For massive galaxies residing in halos of $M_h\approx10^{13}\,\msun$, the vast majority of the mass of the CGM resides in the hot phase, with the cool phase contributing $\lesssim 10$ percent of the total mass of the CGM. This stands in contrast to the CGM mass distribution observed in smaller halos of $10^{12}\,\msun$ (see Fig.\ \ref{fig:phases}), where the cool phase comprises a larger fraction of the total gas mass.  It indicates a shift in the balance between the hot and cool phases as galaxy halos grow in mass.

Spectrally-resolved photoionization analyses offer a range of utilities beyond inferring density and gas mass.  Each best-fit ionization model provides predictions of the ionization fraction of different species.  Combining the observed neutral hydrogen column density $N_{\rm H{\small I}}$ and the anticipated neutral fraction of hydrogen, $f_{{\rm HI}}$, yields a size estimate of the absorbing clump along the line of sight, following $l_{\rm cl}=N_{\rm HI}/(f_{\rm HI}\,n_{\rm H})$.  
Similarly, combining the observed column densities of metal ions and their anticipated ionization fraction leads to estimates of the elemental abundances.  

Most resolved absorbing clumps are found to range in size from approximately 10 pc to 1 kpc, with a mode around $\langle\textit{l}_{\rm cl}\rangle_{\rm mode}=100$ pc, in halos surrounding both star-forming and massive quiescent galaxies \citep[e.g.,][]{Zahedy:2019, Zahedy2021}.  Cool CGM clumps as small as $\textit{l}_{\rm cl}\approx 1$ pc have also been reported \citep[e.g.,][]{Muzahid:2018, Chen:2023b}, revealing fine-grain details of the diffuse gas in distant galaxy halos, similar to what is found in nearby star-forming galaxies (see e.g., Fig.\ \ref{fig:cgmim}).  While uncertainties in the ionizing radiation flux ($\phi$) place a fundamental limit in the inferred density (a higher $\phi$ translates to higher $n_{\rm H}$ and smaller $l_{\rm cl}$; see e.g., \citealt{Zahedy:2019} and Section \ref{sec:abs} above), the clump sizes inferred from the best-fit photoionization models are consistent with those derived for \ion{Mg}{II}-absorbing from comparing the absorption profiles across multiply-lensed quasar sightlines \citep[e.g.,][]{Rauch2002}.  The consistency lends strong support for the robustness of the spectrally resolved photoionization analysis.  Given known gas density, one can compute the Jeans length, which determines the maximum extent over which the clumps are stable against gravitational collapse, following the relation $\textit{l}_{\rm J}\equiv c_s/\sqrt{G\mu m_{\rm H}\,n_{\rm H}}$ \citep[e.g.,][]{Binney:2008}. Here $G$ is the gravitational constant, $c_s$ is the sound speed and $\mu$ is the mean molecular weight.  
For a typical gas density of $n_{\rm H}\sim 0.01\,\cmjjj$, the corresponding Jeans length is $\textit{l}_{\rm J}\approx 20$ kpc, which is significantly larger than the inferred clump sizes.  Therefore, while these cool photoionized clumps are relatively small, thermal pressure is sufficient to support them against gravity.

Similar to gas densities, CGM metallicity---measuring the abundance of elements heavier than helium---also exhibits a large spread along individual sightlines, from less than 1\% solar to super solar values ({\it right} panel of Fig.\ \ref{fig:global}). The large scatter implies that chemical enrichment may be localized and mixing is inefficient \citep[e.g.,][]{Schaye:2007}. Because outflowing materials from star-forming regions are expected to be enriched by metals while newly accreted gas from the IGM is expected to be low in its metal content \citep[e.g.,][]{Peroux2020}, the simultaneous presence of metal-poor and metal-rich gas also supports the notion of a dynamic halo being regulated by inflows and outflows at once (see e.g., Fig.\ \ref{fig:cgmim}).  However, detecting near-solar to super-solar clumps at $d\gtrsim 100$ kpc is surprising.  Comparing the elemental abundance pattern with different enrichment sources may provide key insights into the origin of these highly enriched clumps (see Section \ref{sec:chemical} below).

\subsection{Thermodynamics and energy balance} \label{sec:thermo}

In addition to resolving small-scale density variations (see Fig.\ \ref{fig:global}) and revealing the complex multiphase structure of the CGM (Fig.\ \ref{fig:qal}), high-resolution absorption spectroscopy has enabled detailed investigations into gas kinematics---specifically, the relative motion of gaseous clumps along the line of sight in the rest frame of their host galaxies.  In a multiphase CGM where cool, dense gas clouds are confined by pressure with a more rarified and volume-filling hot gas (Fig. \ref{fig:phases}), the motions of a population of these cool clouds convey critical information about how the different phases of the CGM interact with each other, as well as the physical mechanisms with which such interactions operate.  At a fundamental level, these line-of-sight kinematics provide critical insights into the origin and ultimate fate of the gas.  In turn, the answer informs the overall understanding of how galaxies of varying masses contribute to the enrichment of the IGM and, by extension, the Universe with heavy elements over cosmic time.

For example, under 
hydrostatic equilibrium,
the absorbing clumps are expected to follow the virial motion dictated by the dark matter halo.  The velocity dispersion of these absorbing clumps should align with the halo’s virial velocity, $\varv_{\rm vir}\approx 130\,\kms\,(M_h/10^{12}\,\msun)^{1/3}\,(1+z)^{1/2}$ \citep[e.g.,][]{Mo2010}.  Accordingly, the observed velocity dispersion projected along the line of sight should follow $\sigma_\varv=\varv_{\rm vir}/\sqrt{3}$.  In addition, whether or not the gas remains gravitationally bound to the host halo or escapes into the IGM depends on how deep the gas clumps are in the potential.  This can be evaluated based on the escape velocity $\varv_{\rm esc}(r)=\sqrt{2|\Phi(r)|}$ expected at the distance $r$ from the halo center, which is defined as the velocity necessary to compensate for the gravitational potential $\Phi(r)$ for the gas to escape. Although simplified, comparing the observed line-of-sight velocity offset of the clumps from the host galaxy with the predicted $\varv_{\rm esc}$ for a given $r$ and $\Phi$, typically modeled using the \cite[][NFW]{Navarro:1997} profile, provides an initial assessment for the likelihood of the absorber being unbound.

\begin{figure}[t]
  \centering
  \includegraphics[width=0.825\textwidth]{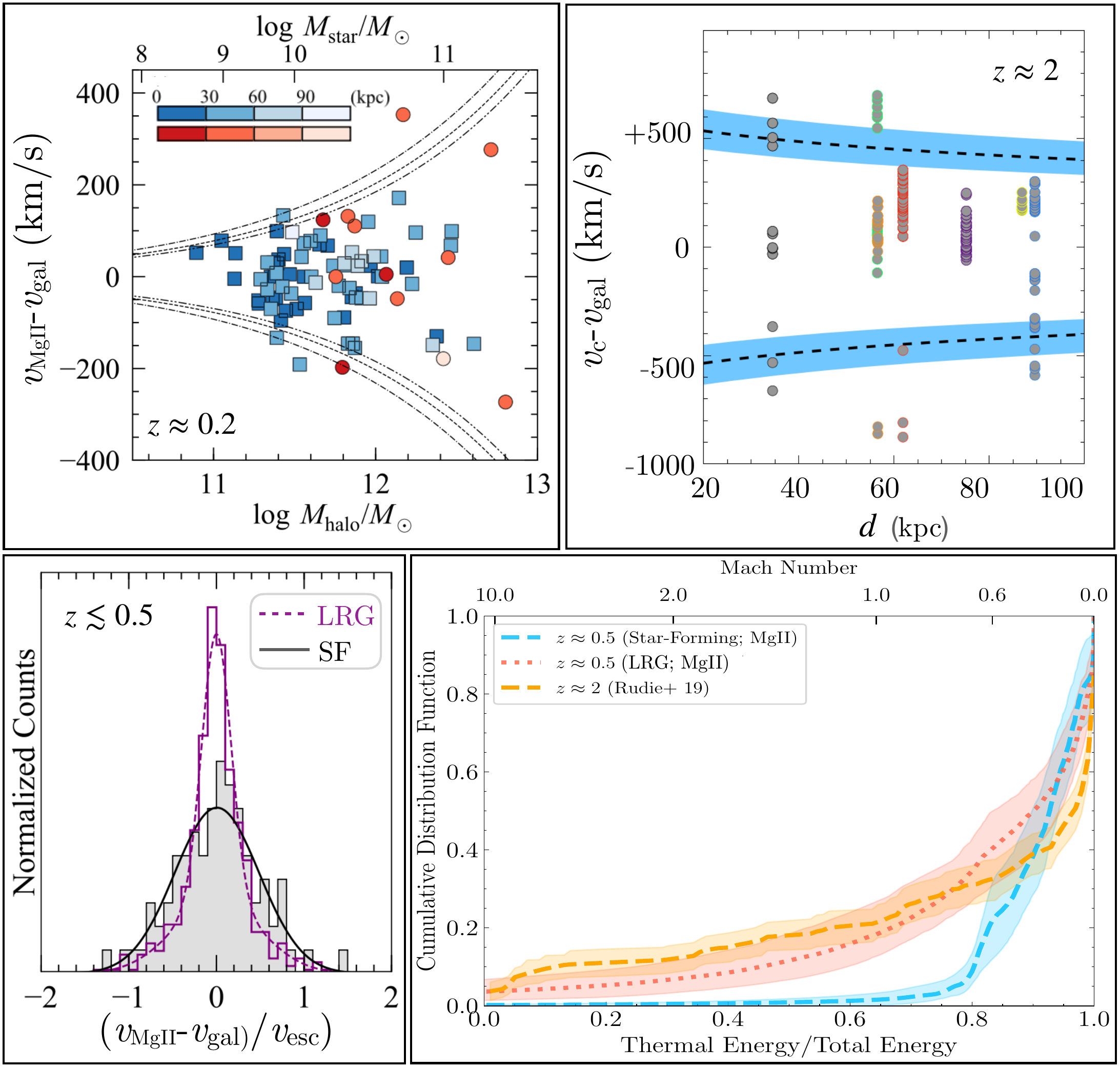}
  \caption{Gas kinematics in the galactic atmosphere revealed in absorption line profiles provide independent clues for the physical origin and fate of the observed cool CGM clouds.  The relative velocity offsets between the systemic velocity of the host galaxy, $\varv_{\rm gal}$, and the column density weighted velocity centroid of the associated \ion{Mg}{II} absorption complex ($\varv_{\rm MgII}$, {\it top-left}) over an ensemble of \ion{Mg}{II}-galaxy pairs at $z\approx 0.25$ show that the majority of these absorbers are consistent with being gravitationally bound to the host galaxies (adapted from \citealt{Huang:2021}). The curves represent the expected escape velocity projected along the line of sight at distances of 30, 60, and 90 kpc.  A similar finding has also been reported for individual metal absorption components ($\varv_c$) detected in the vicinities of star-forming galaxies at $z\approx 2$ ({\it top-right}; adapted from \citealt{Rudie2019}).  These observations demonstrate that the majority of these metal ions do not escape their host halos \citep[cf.,][]{Ho:2021}. Separating galaxies into sub-samples based on their star formation properties has further uncovered intriguing differences in the cloud motions with absorbing clouds in the vicinities of luminous red galaxies (LRGs; massive quiescent galaxies) displaying sub-virial motions \citep[e.g.,][]{Huang:2016}, while those around star-forming (SF) galaxies showing velocity dispersions consistent with virial motions in the host halos ({\it lower-left}; adapted from \citealt{Huang:2021}). At the same time, a non-negligible fraction ($\approx 20$\%) of cool gas identified around LRGs, massive quiescent galaxies, exhibits predominantly supersonic turbulence.  In contrast, the majority of cool clumps identified in star-forming halos exhibit thermally driven random motions ({\it lower-right}; adapted from \citealt{Qu:2022}).  
  }
  \label{fig:kinematics}
\end{figure}

As a starting point, the {\it top-left} panel of Fig.\ \ref{fig:kinematics} displays the line-of-sight velocity offsets between \ion{Mg}{II}-traced CGM absorption complexes and their host galaxies versus host galaxy mass at $z\approx0.25$ \citep[][]{Huang:2021}.  The line-of-sight velocities are compared directly with the expected escape velocity (projected along the line of sight) of the dark-matter halo, and they are evaluated at various distances in the CGM. This comparison reveals that, in the low-redshift CGM, most observed absorbers are consistent with being gravitationally bound to their host halos. Interestingly, 
an analysis of a sample of seven star-forming galaxies at $z\approx 2$ with associated metal absorption components has revealed that five of these galaxies ($\approx 70$\%) exhibit CGM clouds with velocities exceeding the escape speeds of their halos, as shown in the {\it top-right} panel of Fig.\ \ref{fig:kinematics} from \citealt{Rudie2019})

The contrasting CGM kinematics between low- and high-redshift galaxies can be understood by considering two key factors. At higher redshifts ($z\gtrsim 2$), typical galaxies are less massive and experience more intense star formation compared to their low-redshift counterparts. In contrast, galaxies at lower redshifts are more massive and quiescent in star formation, possess deeper potential wells, and drive less powerful outflows, leading to more bound gas in the CGM and less metal enrichment of the surrounding IGM. These differences in mass and star formation rates across cosmic time naturally lead to varying CGM kinematics between high- and low-redshift galaxies.  The expectation of high-redshift low-mass galaxies driving vigorous outflows to push metal-enriched gas out of the galaxies and into the surrounding medium is qualitatively consistent with the observations of IGM being enriched to a mean metallicity of $Z\approx10^{-3}\, Z_\odot$ as early as $z\approx4-5$ \citep[e.g.,][]{Schaye:2003}.

A particularly informative diagnostic for evaluating whether the observed clump motions align with the virial velocities of their host galaxies is the velocity dispersion between absorbers and galaxies over an ensemble of absorber and galaxy pairs.  The {\it bottom-left} panel of Fig.\ \ref{fig:kinematics}) displays histograms of line-of-sight velocity distribution, normalized by the escape velocity $\varv_{\rm esc}$, for \ion{Mg}{II}-traced cool CGM clouds in separate samples of Milky-Way like ($M_h\sim10^{12} M_\odot$) star-forming and more massive ($M_h\sim10^{13} M_\odot$) quiescent (LRG) halos.  Normalizing by the escape velocity allows for a direct comparison of internal halo dynamics across galaxies of widely varying mass, placing them on a comparable scale.  One particularly striking feature in this plot is the significant differences in cloud motions between massive quiescent LRG halos and lower-mass star-forming halos. Cool CGM clouds around high-mass LRGs exhibit sub-virial motion characterized by a velocity dispersion that is only $\approx50-60\%$ of what is expected from virial motion, whereas clouds around lower-mass ($M_h\sim10^{11-12} M_\odot$) star-forming galaxies show a velocity dispersion that is consistent with virial motion expected for their host halos \citep[e.g.,][]{Huang:2021,Anand:2021}. 


The suppressed line-of-sight velocity of cool CGM clouds in high-mass halos has important implications for halo gas dynamics, suggesting that these clouds are affected by physical processes beyond gravitational interactions \citep[e.g.,][]{Huang:2016,Zahedy:2019,Afruni:2019}. Specifically, in the presence of a volume-filling hot CGM, moving cool clouds are subject to a ram-pressure drag force that slows them down \citep[e.g.,][]{Gunn:1972}, until they eventually reach terminal speeds that are less than the virial velocity of the halo. For these dissipative hydrodynamic interactions to slow down the cool gas appreciably on timescales shorter than the orbital time (which is required by the observations), these cool clumps cannot be too massive to begin with, with a mass upper limit that has been estimated to be $\sim10^4 M_\odot$ \citep[e.g.,][]{Zahedy:2019,Afruni:2019}. At the same time, thermal interactions via conduction with the surrounding hot CGM will cause the cool clouds to evaporate over time, as heat is gradually transferred from the hot to the cool phase. For a $\sim10^4 M_\odot$ cool cloud, the expected evaporation time is significantly shorter than the dynamical timescale of the halo \citep[e.g.,][]{Huang:2016}. While these calculations are highly uncertain due to the unknown cloud geometry and the details of thermal conduction, the most likely outcome is that cool clumps in massive LRG halos traverse only a relatively small distance in the CGM during their lifetimes, and most of these clouds will evaporate before reaching the massive galaxies at the center of the halos.

Further insight into the thermodynamic state of the CGM can be obtained by examining gas motions within individual gas clouds themselves, which is afforded by high-resolution spectroscopy from both ground and space. As described in Section \ref{sec:abs}, both thermal and non-thermal motions of atoms and ions in a gas cloud contribute to the observed line broadening. The mass-dependent line broadening from thermal motion and the mass-independent line broadening due to non-thermal motion (e.g., turbulence) can be distinguished by observing two or more ions of different atomic masses. These observations provide a quantitative measure of the internal energetics of these clouds. In particular, energetic processes such as supernova and AGN feedbacks are expected to inject thermal energy into the CGM while also increasing gas turbulence, which may limit the ability of the CGM to infall, replenish the ISM, and fuel future star formation \citep[e.g.,][]{Gaspari2018,Schneider:2020}. 

By decomposing the observed total kinetic energy into thermal and non-thermal contributions, one can compute the thermal energy fraction following the relation
$E_{\rm th}/E_{\rm tot}=b_{\mathrm{th, HI}}^2/(b_{\mathrm{th, HI}}^2+\mu b_{\mathrm{nt}}^2)$, where $\mu=0.6$ is the mean molecular weight for ionized gas \citep[e.g.,][]{Qu:2022}.  It is straightforward to show that the thermal energy fraction is related to the Mach number of the gas, ${\cal M}\equiv \varv_{\rm nt}/c_s$, following $E_{\rm th}/E_{\rm tot}=(1+\gamma(\gamma-1){\cal M}^2/2)^{-1}$ \citep[e.g.,][]{Rudie2019}, where $\gamma$ is the adiabatic index and $\gamma=5/3$ for monatomic gas.  As shown in the {\it bottom-right} panel of Fig.\ \ref{fig:kinematics}, examining the internal energetics of CGM clouds in different galaxy samples has uncovered a striking difference between lower-mass star-forming galaxies and high-mass quiescent LRGs. Specifically, while the majority of cool clumps are subsonic with the velocity field driven by thermal motions, an appreciable fraction of cool gas in the CGM massive quiescent galaxies (LRGs) displays non-thermal and supersonic turbulence. While these results are intriguing, it should be noted that this comparison was performed for galaxy halos of significantly different masses, with LRG halos being on average an order of magnitude more massive than star-forming halos. Despite this caveat, the increased turbulence of cool clouds in LRG halos, in addition to their observed sub-virial motions, provides a tantalizing hint for a lack of star formation in these quiescent galaxies, despite the presence of significant amounts of cool gas in their CGM.

\subsection{Chemical enrichment} \label{sec:chemical}

Observations of diffuse gas in the CGM have traditionally relied on detecting the presence of heavy elements at large distances from galaxies. As discussed in Section \ref{sec:phases} and illustrated in Fig.\ \ref{fig:phases}, various metal ions have been identified, each tracing different density and temperature regimes in the CGM.  The presence of these ions has direct implications for the thermal and chemical states of the gas. For example, the radiative cooling function depends sensitively on the gas metallicity \citep[e.g.,][]{Mo2010}, with higher-metallicity gas cooling more rapidly under the same radiation field for a fixed density. Gas metallicity is also strongly correlated with the total dust content, with higher-metallicity gas showing a higher fraction of dust at a given gas mass \citep[e.g.,][]{decia:2016}. Gas metallicity is, therefore, a key quantity that determines the physical conditions of gaseous clouds in a wide range of astronomical environments \citep[e.g.,][]{Draine2011}.

Since heavy elements are produced by stars in galaxies, gas metallicity also serves as a key measure of a galaxy's overall production of heavy elements. Large-scale spectroscopic surveys of local galaxies have shown that galaxies follow a well-established mass-metallicity relationship, with more massive galaxies exhibiting higher stellar \citep[e.g.,][]{Kirby:2013} and gas-phase \citep[ISM; e.g.,][]{Tremonti:2004} metallicities compared to their lower-mass counterparts. As a result, gas ejected from high-mass central galaxies is expected to be more metal-enriched than gas stripped from lower-mass satellite galaxies. Moreover, recently accreted gas from the IGM should have even lower metallicity, given its chemically primitive nature \citep[e.g.,][]{VanDeVoort:2012, Angles-Alcazar:2017}.

\begin{figure}[t]
  \centering
  \includegraphics[width=0.9\textwidth]{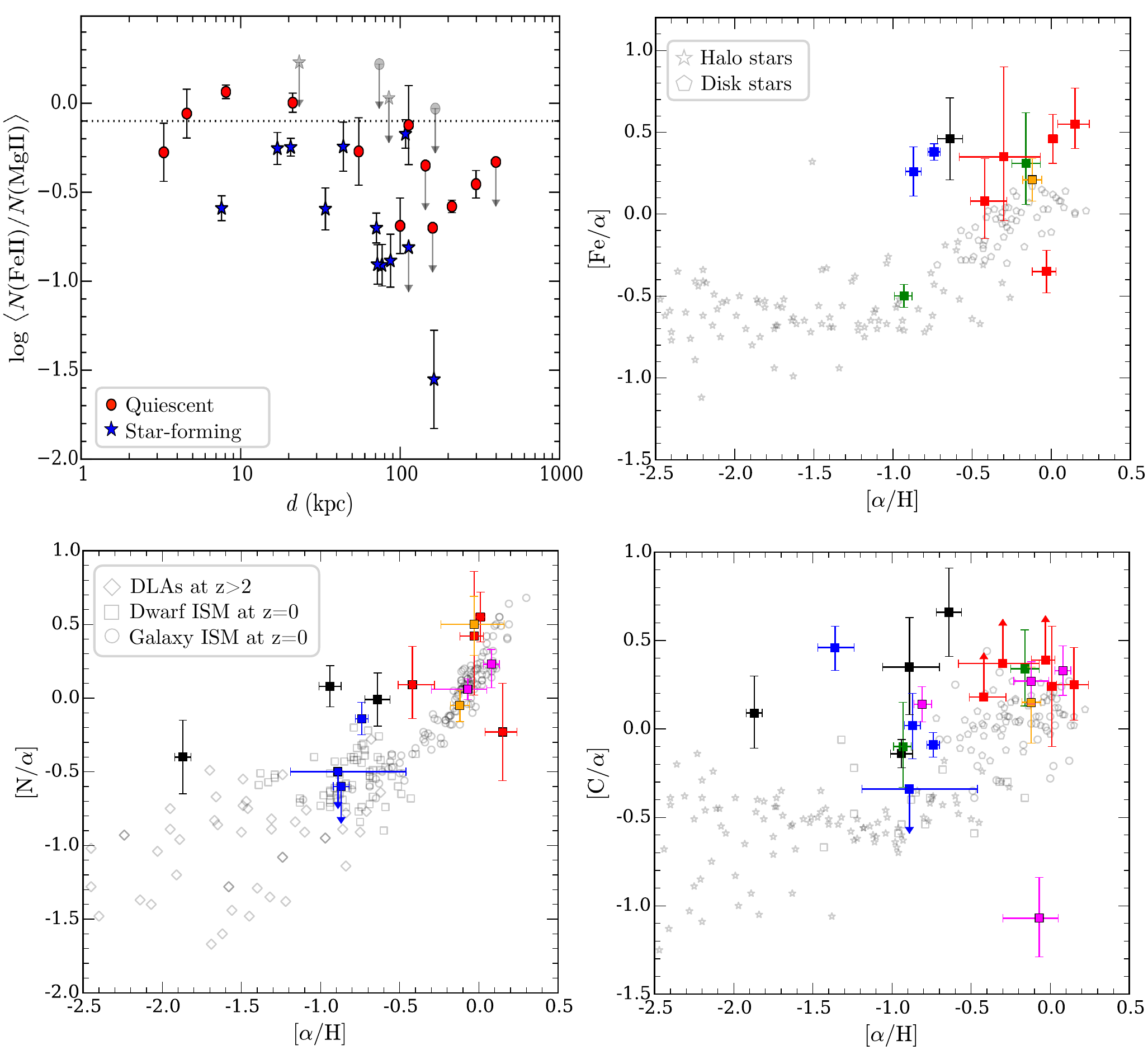}
  \caption{Chemical tagging based on the relative abundances of different elements provides a timing clock for identifying the enrichment sources, thereby constraining the physical origin of the gas. Specifically, the abundance ratios between $\alpha$-elements (e.g., Mg, O), produced promptly over short ($\sim$\,Myr) timescales in massive stars and core-collapse supernovae (SNe), and elements produced in Type Ia SNe (e.g., Fe) and intermediate-mass stars (e.g., N, C) over longer ($\gtrsim0.1$\,Gyr) timescales have long been used to chronicle the chemical enrichment history in Galactic stars and the ISM of other galaxies \citep[e.g.,][]{McWilliam:1997, Weinberg:2019}. In the CGM, these measurements reflect the chemical enrichment sources and timing of heavy element ejection from galaxies.  Indeed, the cool CGM generally displays a declining observed Fe/Mg relative abundance ratio with distance, indicating a chemical enrichment history driven by massive stars in halo outskirts \citep[{\it upper-left} panel][]{Zahedy:2017}.  At the same time, quiescent galaxies display enhanced Fe abundances, comparable to the solar value (dotted line in the {\it upper-left} panel) in the inner halos of quiescent galaxies, suggesting significant contributions from evolved stellar populations.    
  In addition, while the observed spread in metallicity (see Fig.\ \ref{fig:global}) suggests that chemical mixing is inefficient, the observed spread of $\mathrm{[Fe/\alpha]}$, $\mathrm{[N/\alpha]}$, and $\mathrm{[C/\alpha]}$ within a single halo (points sharing a common color in the {\it upper-right} and {\it bottom} panels) indicates that chemical mixing is a complex physical process \citep[e.g.,][]{Zahedy2021,Cooper:2021}. In particular, the presence of a handful of metal-poor gas ($\mathrm{[\alpha/H]}\lesssim-0.1$) with otherwise enhanced Fe, C, and N abundances suggests dilution of stellar ejecta by recently accreted, chemically primitive gas from the IGM. }
  \label{fig:metals}
\end{figure}

However, these straightforward expectations are complicated by the poorly understood chemical mixing processes and heavy element transport in the baryon cycle. If chemical mixing is efficient, with mixing timescales shorter than the dynamical time of the halo, high-metallicity gas expelled from the ISM can rapidly combine with lower-metallicity gas in the outskirts of the CGM, diluting the mixture and reducing its overall metallicity. In such cases, the observed metallicity would no longer reflect the source of the gas in a chemically evolved environment. In contrast, if chemical mixing is highly inefficient, CGM gas could retain its original metallicity long after being ejected from galaxies, leading to significant variations in metallicity within the same galaxy halo. Indeed, as shown in the right panel of Fig.\ \ref{fig:global} and in Fig.\ \ref{fig:metals}, large metallicity spreads---ranging from factors of 10 to 100---are commonly observed in the CGM of individual galaxies across a broad range of masses \citep[e.g.,][]{Zahedy:2019, Zahedy2021, Sameer:2024}, highlighting the complex chemical enrichment history of CGM gas. Therefore, gas metallicity alone is an incomplete diagnostic of the physical origins of CGM gas.

Elemental abundance ratios in gas serve as an archaeological record of the various sources of heavy element production. For instance, $\alpha$-process elements (e.g., O, Mg, Si, and S) are produced through the rapid capture of $\alpha$ particles (helium nuclei), a process that occurs predominantly in the late evolutionary stages of massive stars ($M\gtrsim10\,M_\odot$) and their subsequent core-collapse supernovae (SNe CC). This relatively prompt origin of $\alpha$ elements contrasts with that of Fe-peak elements (e.g., Fe, Ni, Mn), which are produced in large quantities by Type Ia supernovae \citep[SNe Ia][]{Kobayashi:2020}. SNe Ia arise from white dwarf remnants of lower-mass stars ($\lesssim 8\, M_\odot$), whose longer lifetimes delay the release of Fe-peak elements compared to the more rapid enrichment from SNe CC \citep[e.g.,][]{Maoz:2012}. As a result, the observed $\mathrm{[Fe/\alpha]}$ abundance ratio is particularly valuable, as it not only traces the relative contributions of SNe CC and SNe Ia to chemical enrichment but also provides a timing `clock' for the release of heavy elements from galaxies, offering insights into the physical origins of the gas.

The utility of investigating CGM chemistry is illustrated in Fig.\ \ref{fig:metals}. The {\it top-left} panel shows the radial profile of $\mathrm{Fe/Mg}$ elemental abundance ratio in the cool CGM of intermediate-redshift star-forming and quiescent galaxies \citep[][]{Zahedy:2017}, and several noteworthy features are immediately apparent. First, the cool CGM exhibits a general trend of declining $\mathrm{Fe/Mg}$ abundance ratio with increasing distance from galaxies. In the outskirts of the CGM at $d\gtrsim100$ kpc, Mg-rich gas with $\mathrm{[Fe/Mg]<-0.3}$ is relatively common, which is consistent with an early chemical enrichment history that is dominated by massive stars and SNe CC. Secondly, the CGM of quiescent galaxies displays Fe enhancement relative to star-forming galaxies, especially at small projected radii of $d\lesssim10$ kpc, where solar or supersolar $\mathrm{{Fe/Mg}}$ values are commonly observed. The observed Fe enhancement can be understood as a result of significant contributions from evolved stellar populations in quiescent galaxies, leading to a higher proportion of SNe Ia than seen in the solar neighborhood \citep[][]{Zahedy2016, Zahedy:2017}. 

While the large spread in CGM metallicity shown in Fig.\ \ref{fig:global} indicates that chemical mixing in the CGM is an inefficient process, further insight into chemical mixing can be obtained by examining various elemental abundance ratios in the CGM as a function of gas metallicity. This exercise is illustrated in the remaining three panels of Fig.\ \ref{fig:metals}, which displays the observed $\mathrm{[Fe/\alpha]}$, $\mathrm{[N/\alpha]}$, and $\mathrm{[C/\alpha]}$ abundance ratios in the CGM as a function of metallicity $\mathrm{[\alpha/H]}$, in comparison with existing constraints on these quantities for Galactic stars as well as ISM gas at low and high redshifts \citep[][]{Zahedy2021,Cooper:2021}. In contrast to $\alpha$-process elements primarily produced in massive stars and SNe CC and Fe-group elements produced largely in SNe Ia, nitrogen and carbon owe their existence largely from intermediate-mass asymptotic giant branch (AGB) stars \citep[e.g.,][]{Romano:2022}. 

The complex nature of chemical enrichment in the CGM is evident in the substantial spread ($>1$ dex) in both the absolute gas metallicity $\mathrm{[\alpha/H]}$ and relative elemental abundance ratios. At the same time, investigations of the observed variations in $\mathrm{[\alpha/H]}$, $\mathrm{[Fe/\alpha]}$, $\mathrm{[N/\alpha]}$, and $\mathrm{[C/\alpha]}$ within {\it individual} galaxy halos are also instructive. First, a clear majority ($\gtrsim 60\%$) of detected CGM clouds have chemical properties (metallicity and elemental abundance ratios) consistent, to within the measurement uncertainties, with the observed metallicities and abundance ratios in stars and the ISM. Taken at face value, a lack of dilution would be a natural consequence of inefficient chemical mixing in the CGM. On the other hand, the presence of a handful of metal-poor gas clouds ($\mathrm{[\alpha/H]}\lesssim-0.1$) with otherwise elevated Fe, N, and C abundances suggests the dilution of chemically evolved gas (stellar ejecta or ISM gas) by more recently accreted gas from the chemically more primitive IGM. 
While this type of detailed chemical study has been performed for a relatively small number of diffuse CGM clouds to date, these investigations highlight the power of using CGM chemistry to resolve the physical origins of halo gas observed at large distances from galaxies. 


\section{Future outlook} \label{sec:future}


Observational studies have revealed the CGM as part of a dynamic, complex, and multiphase ecosystem.  Theory predicts that this gas reservoir plays a crucial role in regulating star formation and chemical enrichment in galaxies across cosmic time \citep[e.g.,][]{Somerville:2015, Naab:2017, Crain:2023}.  
Despite significant advancements in our empirical understanding of the CGM’s global content and its physical and chemical properties, many fundamental questions remain.  These open questions highlight the missing physics in our current understanding of the origin of the multiphase gas, its role in regulating the cosmic baryon cycle, and its connection to galaxy growth and evolution over time. 

Using high-resolution absorption spectroscopy, astronomers have been able to resolve detailed density, thermodynamic, and chemical enrichment structures over a large dynamic range in spatial scales from $\approx 1$ pc to $\approx 1$ kpc in the diffuse CGM to density as low as $n_{\rm H}\sim 10^{-4}\,\cmjjj$ (Section \ref{sec:abs}).  However, these studies have largely been limited to one-dimensional (1D) probes, making establishing direct connections to galaxy properties along transverse directions to the absorbers challenging.  On the other hand, detections of line-emitting signals from the diffuse CGM provide two-dimensional (2D) morphologies for establishing direct connections between star formation and AGN activities in galaxies and the extent, kinematics, and ionization conditions of the gas (Section \ref{sec:emission}).  However, emission measures are biased toward the highest-density clumps and severely affected by the $(1+z)^4$ cosmological dimming.  Current emission observations are largely limited to detecting \lya\ photons, which are challenging to interpret due to their resonant scattering nature.

\begin{figure}[t]
  \centering
  \includegraphics[width=0.9\textwidth]{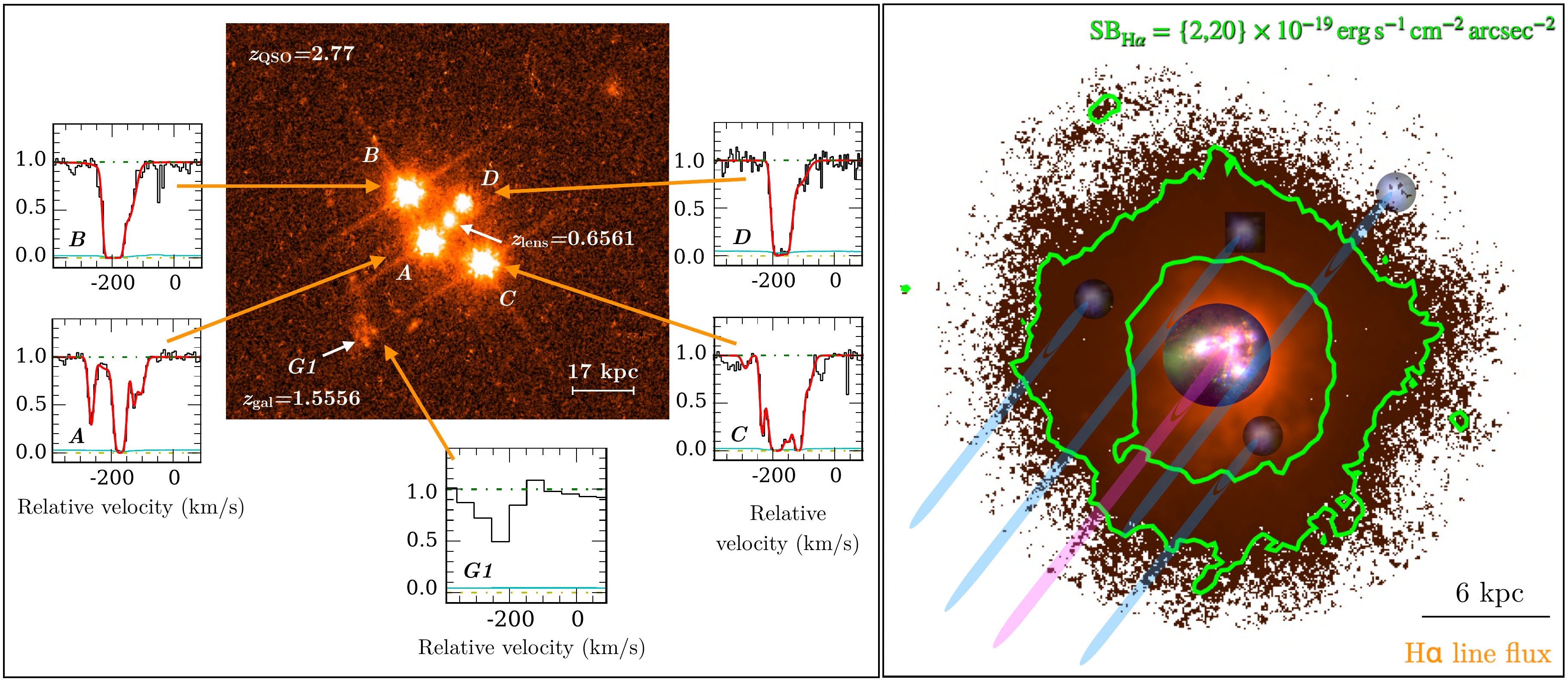}
  \caption{Mapping the multi-scale and multiphase galactic atmosphere in absorption {\it and} emission. The {\it left} panel showcases an example of spatially and spectral resolving gas flows in the CGM of a star-forming galaxy at $z_\mathrm{gal}=1.556$, using a combination of a quadruply-lensed quasar at $z_\mathrm{QSO}=2.77$ and stars in the galaxy itself (Zahedy et al.\ in preparation).  The presence of a blueshifted absorption profile in the galaxy's spectrum ($G1$ panel) confirms that the galaxy is actively driving outflows (zero velocity corresponds to the galaxy's systemic redshift determined from [\ion{O}{II}] emission lines), while the four lensed quasar sightlines ($A$ to $D$) probe the galaxy's CGM as projected distances between $d=39$ and $d=45$ kpc.  The high spectral resolving power afforded by the lensed quasar has helped reveal chaotic gas motions with velocity dispersion exceeding 100 \kms\ across all four sightlines.  The {\it right} panel showcases a spatially extended H$\alpha$ line emitting halo uncovered around a low-mass starburst galaxy with sensitive IFS observations.  The size of the H$\alpha$ halo is $\approx 5$ times larger than the stellar body, revealing remnants of tidal interactions that triggered the starburst phase \citep[see also][]{Menacho:2019}.  Because the emission signals scale with gas density according to $n_{\rm H}^2$, the intensity map is most sensitive to high-density peaks.  Applying absorption spectroscopy of multiple background sources with future large telescopes will enable a dense sampling of the low-density gas missed by the emission-line maps.
  } 
  \label{fig:future}
\end{figure}

\subsection{Spatially and spectrally resolve the multiscale and multiphase CGM in absorption and emission} \label{sec:multisightlines}

Spatially and spectrally resolved maps of the CGM around galaxies of different star formation history and over a broad redshift range hold the promise to resolve the complex multiscale physics and establish a direct connection between gas and stars in the cosmic ecosystems.  This can be achieved with upcoming surveys using several different techniques, including emission observations using high-throughput wide-area integral field spectrographs (IFSs) on large ground-based telescopes and high-resolution absorption spectroscopy of multiply-lensed quasars uncovered from deep all-sky surveys.  Combining 2D emission maps with high-resolution absorption spectra of multiply-lensed background objects enables a full exploration of this cosmic gas reservoir based on a high-definition 3D view of its density, velocity, and chemical enrichment structures.  

Luminous quasars, multiply lensed by a foreground object, expand traditional quasar absorption spectroscopy from 1D to 2D probes, and provide additional spatial resolving power for constraining the size and velocity (de)coherence of absorbing gas \citep[e.g.,][]{Rauch2001, Zahedy2016}.  A gravitationally lensed QSO occurs when the strong gravitational field of a massive galaxy situated directly between the observer and a distant QSO produces multiple lines of sight corresponding to the multiple images of the quasar, as shown in the {\it left} panel of Fig.\ \ref{fig:future}. As a result of this serendipitous alignment, the multiple images (usually two to four) of a lensed QSO act as independent probes to resolve the CGM and represent a powerful means to measure the kinematic and chemical coherence scales in the CGM and identify different gas flows and constrain their properties based on the additional spatial information.

The example displayed in Fig.\ \ref{fig:future} showcases a quadruply-lensed QSO at $z_\mathrm{QSO}=2.77$ and a star-forming galaxy $G1$ at $z_\mathrm{gal}=1.556$. This foreground galaxy $G1$ is driving a strong outflow, as evidenced by the presence of a blueshifted absorption profile in its spectrum. The four lensed lines of sight ($A$ to $D$) each probes the gaseous halo of the galaxy at projected distances of between 39 and 45 kpc (Zahedy et al.\ in preparation). High-resolution absorption spectra of these sightlines all show the abundant presence of cool gas around this galaxy. At the same time, significant differences are seen in the absorption profiles among four sightlines separated by only a few kpc. Specifically, while the gas exhibits kinematically complex with large velocity spreads ($\gtrsim200 \,\kms$) along sightlines $A$ and $C$ at $d\approx40$ kpc, the observed kinematics are comparatively simpler with smaller velocity spreads along $B$ and $D$ probing distances a few kpc farther from $G1$. These observations suggest that the outflowing material from $G1$ either decelerates and/or becomes more disrupted and less well-collimated with increasing distance.  A quantitative analysis will provide strong constraints on the outflow geometry and energetics, and investigating the observed variations in gas chemistry across the four sightlines will provide critical insight into the physical origins of the gas.  

Since the discovery of the first doubly imaged lensed quasar in 1979 \citep[][]{Walsh:1979}, nearly 300 have been discovered to date \citep[][]{Lemon:2024}, and around a few dozen are bright enough to obtain high-resolution spectra of with current facilities. The upcoming all-sky Legacy Survey of Space and Time (LSST) at the Rubin Observatory is expected to discover approximately $8000$ new lensed quasars \citep[e.g.,][]{Oguri:2010} during the survey's duration. Among these, at least 500 will be bright enough for high-resolution observations with future Extremely Large Telescopes (ELTs), representing a more than tenfold increase over the current state of the field. 

In parallel, IFS observations, targeting non-resonant emission lines to a surface brightness limit of $\approx 10^{-19}\,{\rm erg}\,{\rm s}^{-1}\,{\rm cm}^{-2}\,{\rm arcsec}^{-2}$ or fainter, will begin to probe the density and velocity structures in diffuse halo gas around typical star-forming galaxies \citep[e.g.,][]{Corlies:2020,Dutta:2024}.  The {\it right} panel of Fig.\ \ref{fig:future} showcases extended H$\alpha$ line emitting signals around Haro 11, a starburst galaxy at $z\approx 0.02$ \citep[e.g.,][]{Grimes:2009}, uncovered in moderately deep IFS observations.  The line-emitting gas covers an area that is five times the size of the stellar body \citep[see also][]{Menacho:2019}.  The gas is detected in multiple nebular lines, placing strong constraints on the gas density and ionization conditions.  It is found that the gas is primarily photoionized by the young star clusters with a relatively high density in the immediate proximity of the central star clusters with $n_{\rm H}\approx 100\,\cmjjj$.  The gas density declines toward the outskirts with a loose constraint of $n_{\rm H}<40\,\cmjjj$.  The emission morphology shows that the gas is clumpy and reveals clear tidal features extending outward from the nuclear star clusters.

\begin{figure}[t]
  \centering
  \includegraphics[width=0.975\textwidth]{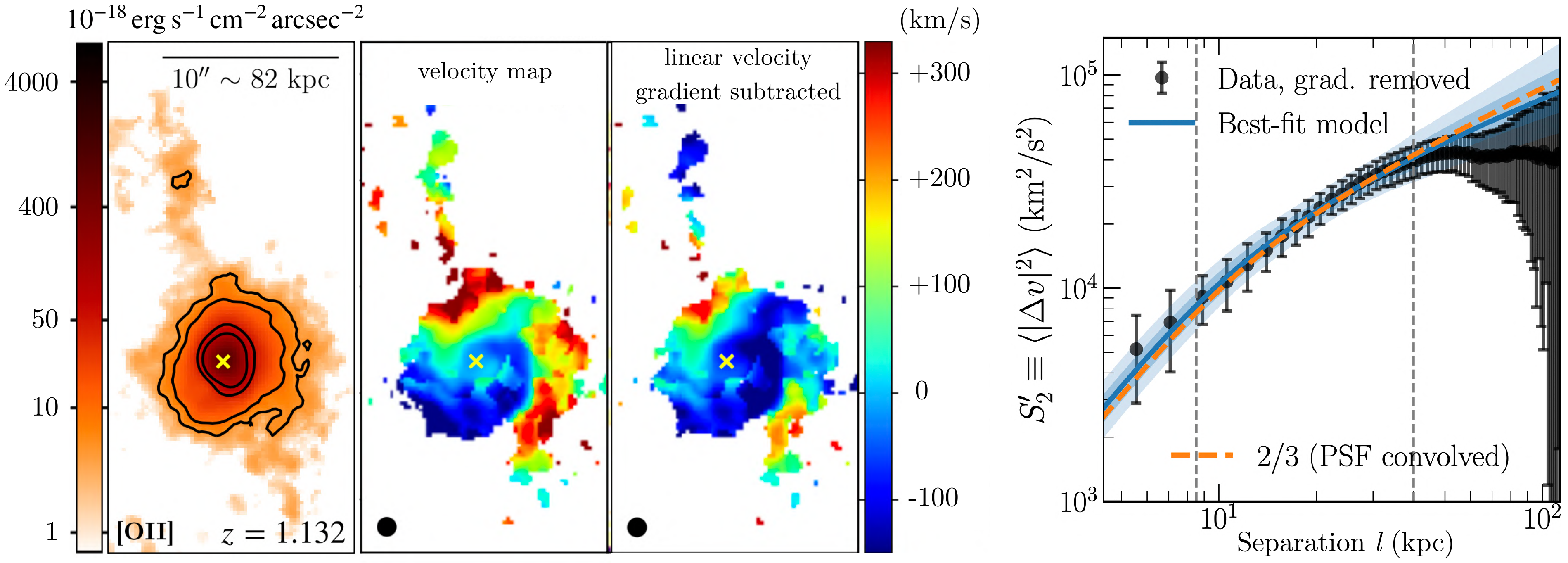}
  \caption{Turbulent energy transfer across different spatial scales from $\gtrsim 50$ kpc to $\lesssim 10$ kpc in a giant quasar nebulae \citep[adapted from][]{Chen2023}. 
    The three {\it left} panels display the 2D maps of [\ion{O}{II}] surface brightness, observed line-of-sight velocities, and residual velocities after a best-fit linear gradient is subtracted \citep[from left to right; see also][]{Johnson2022}.  This central [\ion{O}{II}]-emitting nebula with surface brightness exceeding $10^{-18}\,{\rm erg}\,{\rm s}^{-1}\,{\rm cm}^{-2}\,{\rm arcsec}^{-2}$ spans nearly 100 kpc in diameter with fluctuating line-of-sight velocities ranging from $<-100$ \kms\ to $>+300$ \kms\ from the systemic redshift of the quasar at $z_{\rm QSO}=1.132$.  The {\it right} panel displays the residual velocity variance after subtracting a linear velocity gradient, $\langle\,|\Delta\,\varv(l)|^2\,\rangle$, versus projected separation $l$, representing the 2nd-order velocity structure function (VSF).  Accounting for the atmospheric seeing effects that smooth ground-based data and progressively suppress noise at smaller scales, the best-fit model (blue solid curve) aligns well with the Kolmogorov expectation (orange dashed curve), which predicts a slope of $2/3$ for isotropic, homogeneous, and incompressible fluids.  This agreement suggests subsonic turbulence and a constant energy transfer across spatial scales within the quasar nebula.
  } 
  \label{fig:vsf}
\end{figure}

A promising new avenue these spatially resolved maps enable is investigating turbulent energy transfer across spatial scales. This approach offers an empirical framework to understand how feedback energy from massive stars and SMBHs on parsec scales couples to diffuse gaseous halos extending over 100 kpc or more, influencing their thermal state. Specifically, energetic processes associated with stellar activity, SMBHs, and satellite accretion events are expected to stir the surrounding medium, disrupting bulk flows and driving turbulence (see e.g., Fig.\ \ref{fig:cgmim}).  In addition, a cool, ionized plasma of density $n_{\rm H}\!\approx\!0.01\,\cmjjj$ and temperature $T\!\approx\!10^4$ K (see Section \ref{sec:phases}) is expected to have an effective mean free path of $\sim\!10^{14}$ cm and kinematic viscosity of $\approx\!10^{20}\,{\rm cm}^2\,{\rm s}^{-1}$ \citep[e.g.,][]{Spitzer:1962, Sarazin:1986}.  For clumps of size $100$ pc \citep[e.g.][]{Zahedy2021}, moving at speeds of $100$ \kms\ in galaxy halos \citep[e.g.,][]{Huang:2021}, the associated Reynolds number is large, $Re\sim 3\times 10^7$ \citep[cf.][for neutral medium observed in Milky Way high-velocity clouds]{Marchal:2021}. This high Reynolds number suggests that the cool CGM clouds are likely turbulent \citep[see also][]{Burkhart:2021}.

A classical approach in fluid dynamics for characterizing the degree of this decorrelation is to measure the velocity structure functions (VSFs), which are moments of velocity fluctuations ($S_n$) as a function of spatial scales $l$ for quantifying the velocity power spectrum \citep[e.g.,][]{Frisch1995,Boldyrev2002}.  In particular, the second-order VSF, $S_2(l)$, represents the scale-dependent variance of the turbulent medium $|\Delta\,\varv(l)|^2$, which translates directly to the turbulent energy on different scales, and  
the energy transfer between different scales is expected to be conserved in a homogeneous, isotropic, and incompressible fluid.  From dimensional analysis, it is straightforward to demonstrate that a constant energy transfer rate would lead to a constant $\Delta\,\varv^2(l)\times\Delta\,\varv(l)/l$. 
In turn, it gives a power-law relation of $\langle\,|\Delta\,\varv(l)|^2\,\rangle\!\propto\!l^{2/3}$ and an energy transfer rate per unit mass of $\epsilon\!=\!(5/4)\langle\,|\Delta\,\varv(l)|^3\,\rangle/l$ \citep[e.g.,][]{Kolmogorov1941}. 
For subsonic turbulence, kinetic energy injected on large scales is expected to propagate to small scales at a constant rate ($\epsilon$), eventually dissipating at the smallest scale when viscosity transforms the kinetic energy into heat. 
Observations of scale-dependent velocity dispersion, therefore, place direct constraints on turbulent heating in the diffuse CGM.

By targeting multiply-lensed QSOs, \cite{Rauch2001} presented the first VSF measurements of the intergalactic medium at $z\!\approx\!3$ on spatial scales between 30 pc and 30 kpc.  However, the measurements were limited by a small sample of \ion{C}{IV} absorbers found along multiply-lensed QSOs, which led to large uncertainties.  Recent IFS observations of spatially extended line-emitting nebulae have reinvigorated interest in investigating turbulence within the diffuse halo gas through VSF measurements \citep[e.g.,][]{Li2020, Chen2023}.  Initial results have generated new surprises and puzzles.  

Fig.\ \ref{fig:vsf} showcases a giant luminous quasar nebula at $z_{\rm QSO}=1.132$ with [\ion{O}{II}] surface brightness exceeding $10^{-18}\,{\rm erg}\,{\rm s}^{-1}\,{\rm cm}^{-2}\,{\rm arcsec}^{-2}$ and a central body spanning nearly 100 kpc in diameter.  While the velocity map exhibits apparent organized motions in different parts of the nebula, no single coherent flow model, either a rotational, radial, or linear velocity field, can capture these features at once.  The 2nd-order VSF determined based on residual velocities after a best-fit linear velocity gradient is subtracted across the nebula exhibits a spectacular agreement with the Kolmogorov expectation, implying subsonic turbulence and a constant energy transfer rate over the spatial scales covered by the measurements \citep[][]{Chen2023}.  The surprising agreement between the observed $S_2(l)$ and the Kolmogorov expectation has profound implications for the nature of the quasar nebula.  First, subsonic turbulence is at odds with the observed large velocity fluctuations, greater than 100 \kms, exceeding the sound speed $c_s\!\approx\!10$ \kms\ expected for cool, photoionized gas of $T\!\sim\!10^4$ K.  Such discrepancy suggests that the gas is condensed out of the ambient hot halo of $T\!\gtrsim\!10^6$ K inheriting the turbulent energy of the hot halo, rather than being driven out by fast outflows from the central SMBH.  Secondly, the observed turbulent power on the scale of 50 kpc implies a turbulent heating time scale of $\approx 200$--300 Myr, which is much longer than the lifetime of the active quasar phase ($<\!10$ Myr) \citep[e.g.,][]{Schawinski2015, Sun2017}.  In contrast, many of these quasars reside in an overdense environment with numerous gas-rich satellites found and a correlation between the presence of close companions and the
presence of strong, extended nebular line emission has been reported \citep[e.g.,][]{Stockton1987, Johnson:2024}.  The dynamical time of satellites orbiting in quasar host halos is comparable to the inferred turbulent heating time scale, making satellite interactions a likely turbulence driver in quasar nebulae.

The example presented in Fig.\ \ref{fig:vsf} illustrates the potential of VSF measurements to place crucial constraints on the turbulent nature of the CGM, offering illuminating insights into the physics of feedback processes. At the same time, it also highlights the limitation both in terms of the detectability of extended line-emitting nebulae and the dynamic range in spatial scales over which the VSF slopes can be robustly measured. As noted in Section \ref{sec:emission}, line-emission signals from the diffuse gas are highly dependent on gas density, with the intensity scaling as $n_{\rm H}^2$. This means that emission-line maps primarily highlight regions of high-density gas, making it challenging to capture low-density areas. However, future large telescopes present a promising opportunity to address this limitation.  Meanwhile, absorption spectroscopy provides a valuable tool to expand the reach both to lower-mass, non-quasar host halos and to spatial scales as small as $\sim 1$ pc \citep[e.g.,][]{Chen:2023b}.  In addition, astronomers can densely sample and detect lower-density gas that emission-line maps may miss by employing absorption spectroscopy on multiple bright background sources \citep[e.g.,][]{Newman:2019}.  Based on the galaxy number counts from deep surveys \citep[e.g.,][]{Metcalfe:2001}, it is expected that there will be on average two background sources brighter than $B\approx 24$ mag and four such sources brighter than $B\approx 24.5$ mag per square arcminute to enable such multiple absorption probes to complement emission measures. Combining emission line maps and multiple absorption line probes will provide spatially and spectrally resolved maps of the multiphase gas to constrain the extent, geometry, and energetics of galactic outflows and accretion over a wide range of galaxy masses and much of cosmic evolution. 

\subsection{Independent constraints from Sunyaev-Zel'dovich signals and fast radio bursts} \label{sec:others}

Moving beyond UV and optical wavelengths, radio and sub-millimeter observations offer exciting avenues for studying the cosmic baryon content.  One example is through the Sunyaev-Zel'dovich (SZ) effect \citep{Sunyaev:1980}, resulting from the scattering of cosmic microwave background (CMB) photons by high-energy electrons in the intracluster medium (ICM) of galaxy clusters. This inverse Compton scattering boosts the energy of the CMB photons by $\approx k_BT_e/m_e c^2$, shifting the CMB spectrum in a way that decreases intensity at frequencies below 218 GHz and increases intensity at higher frequencies.  The anticipated signal strength expressed in terms of the CMB blackbody temperature is proportional to the thermal pressure of the ionized gas integrated along the line of sight, $\Delta\,T_{\rm SZ}/T_{\rm CMB}\propto \int n_e (k_BT_e/m_e c^2)\,\sigma_T\,dl$, where $\sigma_T$ is the Thomson scattering cross-section.
Unlike X-ray observations, which place stronger weights toward denser regions of the ICM, the SZ signals provide a more unbiased measure of ionized gas pressure and offer a means of tracing the baryonic content of the universe throughout cosmic history.  The redshift insensitivity of SZ signals presents both advantages and disadvantages. On the positive side, this property allows SZ observations to probe sources at any distance, unrestricted to nearby objects. However, the downside is that SZ signals alone lack the discriminative capability to investigate the cosmic evolution of ionized plasma.  Such limitations can be mitigated by cross-correlating the SZ signals with galaxies identified in all-sky spectroscopic surveys \citep[e.g.,][]{Chiang:2020}.

As the detector technology advances, CMB telescopes become increasingly sensitive to detect the SZ signals beyond the ICM and into the intragroup gas of lower mass galaxy groups. This makes the SZ effect a powerful tool for probing the diffuse ionized gas associated with overdense galaxy environments. Although the angular resolution of SZ observations remains poor, $\gtrsim 1'$, these data still provide sensitive constraints for the pressure profile beyond the nominal size of massive halos \citep[e.g.,][]{Schaan:2021}.  The Cosmic Microwave Background Stage Four (CMB-S4) experiment, when completed, will have much-improved sensitivities for probing ionized gas associated with lower mass halos, approaching the Milky-Way halo scale \citep{CMBS4:2019}.  These data will enable direct comparisons between CGM observations from different techniques and provide the most stringent constraints on theoretical models.

Independently, fast radio bursts \citep[FRBs][]{FRB:2022} offer a unique tool for probing the cosmic baryon content. These brief, intense flashes of radio waves originating from distant galaxies are dispersed as they travel through an ionized medium, resulting in a delay in the signal's arrival time.  The time it takes for the pulse to arrive is frequency dependent, following the relation $t_{\rm arrival}\propto (\nu/{\rm GHz})^{-2}\,({\rm DM}/{\rm cm}^{-3}\,{\rm pc})$.  Here DM is the dispersion measure, an integral of electron density along the path to the observer, ${\rm DM}=\int n_e\,dl$. This provides a direct probe of the diffuse ionized gas, that is difficult to detect by other means.  While each DM measurement includes all electrons integrated along the line of sight, from the host ISM and CGM of the FRB itself to the CGM associated with intervening galactic halos and the unbound IGM, and to the Milky Way ISM, it is possible to separate the contributions of the IGM and CGM from the host and the Milky Way by incorporating available constraints for the Milky Way ISM and the FRB host \citep[e.g.,][]{Macquart:2020}.  With several radio projects underway globally, such as CHIME \citep{CHIME:2021}, ASKAP \citep{ASKAP:2020}, and HIRAX \citep{HIRAX:2016}, FRB DM studies hold significant promise for measuring cosmic mean baryon content.

\begin{ack}[Acknowledgments]%

The authors express their deep appreciation for the many insightful discussions with colleagues that contributed to shaping the ideas presented in this article. Special thanks are extended to Michael Rauch and John Mulchaey for their invaluable guidance, inspiration, and thoughtful input throughout the authors' careers, and to Zhijie Qu for helpful comments on an earlier draft of this article and for his assistance in preparing the mass budget panel in Figure 2 and the column density profiles in Figure 6. The authors would also like to thank Suoqing Ji for providing updated simulation predictions of the column density profiles in Figure 6.
  
\end{ack}

\seealso{Advanced readers are referred to Saas-fee lecture notes by \cite{Fumagalli:2024} and comprehensive reviews by \cite{Donahue2022, FaucherGiguere2023} for detailed descriptions of relevant multi-scale physical processes that regulate the observed CGM properties.  Earlier reviews on this subject include \cite{Tumlinson2017, Peroux2020}.}

\bibliographystyle{Harvard}
\bibliography{references}

\end{document}